%% file: main.tex
\documentclass {eptcs}
\providecommand{\event}{SCSS 2012} 
 \usepackage{breakurl}       

\usepackage[pdftex]{graphicx}
\usepackage{todonotes}
\usepackage[latin1]{inputenc}

\usepackage[curve]{xypic}
\usepackage{latexsym} 

\usepackage{alltt}
\usepackage{isabelle,isabellesym}
\isabellestyleit

\newcommand{\secref}[1]{Section~\ref{sec:#1}}
\newcommand{\secrefs}[1]{Sections~\ref{#1}}

\newcommand{\tabref}[1]{Table~\ref{tab:#1}}

\title{Formal verification of a proof procedure for the description logic \large {$\cal{ALC}$}}

\author {Mohamed Chaabani \qquad Mohamed Mezghiche
  \institute{LIMOSE, University of  Boumerdes\\
     Boumerdes, Algeria}
  \email{chaabani@umbb.dz \quad\qquad mohamed.mezghiche@yahoo.fr}
   \and
  Martin Strecker
  \institute {IRIT (Institut de Recherche en Informatique de Toulouse)\\
  Université de  Toulouse }
  \email{strecker@irit.fr}
}

\def\titlerunning{Formal verification of a proof procedure for the description logic \large {$\cal{ALC}$}}
\def\authorrunning{M. Chaabani, M. Mezghiche \& M. Strecker}

\begin{document}
\maketitle

\begin{abstract}
Description Logics (DLs) are a family of languages used for the
  representation  and for reasoning about the knowledge base of an application
  domain, in a structured and formal manner.To achieve this objectif, sevral provers have been implemented such as RACER and FACT++, but these
  provers themselves have not been certified. In order to insure the soundness
  of derivations in these DLs, it is necessary to verify formally the
  deductions applied by these reasoners. Formal Methods offer powerful tools
  for the specification and verification of  proof procedures,
  among them methods of proving properties such as soundness, completeness and
  termination of a proof procedure.

  In this paper, we present the definition of a proof procedure for the
  Description Logic $\mathcal{ALC}$, based on a tableau method. 
  We then prove the soundness, completeness and termination of our reasoner
  with the proof assistant Isabelle. The
  proof proceeds in two phases, by first establishing these properties on an
  abstract, set theoretic level, and by instantiating them with an
  implementation based on lists.
\end{abstract}

\input{intro}                   
\input{DLs}                     
\input{Syntax}                  
\input{Semantic}                
\input{Tableau}                 
\input{Soundness}               
\input{Completeness}            
\input{Implementation}          
\input{conclusion}              

\bibliographystyle {eptcs}
\bibliography{main}

\end{document}

%% file: intro.tex
\section{Introduction}\label{sec:introduction}

In this paper, we present a definition of a prover for the description logic $
\mathcal {ALC} $ \cite{Schmidt-SchaubB:1991:ACD:114341.114342} which is based on the method of semantic tableau \cite{Baader00tableaualgorithms}. We ensure
the validity of our method and its implementation by the proof of the
properties of its soundness, its completeness and its termination. These proofs
are performed using the Isabelle/HOL proof assistant.

Description Logics (DL) are formalisms  widely used in several areas such as the
Semantic Web and ontology construction. Several description logics, eg,  
$\mathcal {SHOIQ} $ \cite {Horrocks} and $\mathcal {SHOIN}$
\cite{BaaderHS05} are more expressive extensions of $ \mathcal {ALC} $.
$\mathcal {SHOIN} $, which is considered as a formal model of Web Ontology
Language OWL-DL, is used in several provers such as FaCT++
\cite{Tsarkov2006} and  RACER \cite{Racer} .

Only a formal proof of the validity of the reasoning process applied to DLs
can ensure the correctness of derivations of properties in these logics. This
is far from being the case for provers available today. The inference engines of provers like
RACER and FaCT++ have not yet been certified. In \cite{luther09:_who_heck_father_bob}, the authors give a detailed description of the
various problems posed by incomplete and incorrect provers yet widely used.

Of the many methods used as decision procedures for DLs, the  semantic tableau method is
the most common. Indeed, FaCT and RACER use it in their reasoning process. We
propose in this paper a definition of a prover based on a semantic tableau
method for the description logic $ \mathcal {ALC} $. We will also present the
formal proof in the Isabelle/HOL proof assistant of soundness,
completeness and termination of the prover.

Work close to our formalization of proof procedures are reported in  \cite{ridge05:_fol},
\cite{wind01:_modal_logic} and \cite{schimpf:construction}. The first describes
the formalization of a prover for  first-order logic in the
Isabelle/HOL proof assistant. The second presents a formalization in the Coq
proof assistant of some modal logics, and the third a formalization of a
prover for LTL. Please note that DLs can be considered as specific modal
logics. Closest to  our work comes  \cite {hidalgo07:_alc}, which describes the formalization of $
\mathcal {ALC} $ in the PVS proof assistant. Our formalization simplifies the
termination argument, see \secref {implantation}, and allows to extract a prover  directly
executable in a `` standard'' programming language (Caml). This paper builds on and extends our previous work described in \cite{chaabani09:_formal_de_la_logiq_descr}, which provides an  abstract, non-executable proof procedure for $ \mathcal {ALC}$, that extension  is presented in \cite{chaabani10_afadl}.

This paper is organized as follows: In \secrefs{sec:log_descr}, \ref {sec:syntax} and \ref{sec:semantique}, we detail the description logics. In
\secref {tableau}, we present the procedure of semantic tableau for $ \cal {ALC}$
and the \secrefs {sec:correction} and \ref {sec:completude}, we
describe the formalization of this method  and present the proof of soundness and completeness properties in the Isabelle/HOL proof assistant. In
the rest of the article, we present an implementation of a method of proof for
the logic $\mathcal {ALC}$ and the proof of its soundness and its
termination.

The development described here \footnote{http://www.irit.fr/CMS-DRUPAL7/AcadieProjects/dl\_verified} was carried out in the environment of the
Isabelle/HOL proof assistant \cite {Isabelle_Tutorial} whose logic HOL is a
classical logic. However, the background of the development is largely independent of Isabelle
and could easily be simulated in other proof assistants.


%% file: DLs.tex
\section{ Description logics }\label{sec:log_descr}
Description Logics
\cite{dlhandbook,Baader1991,BaaderSattler-JLC-99,DBLP:conf/kr/DoniniLNN91} are
a family of knowledge representation languages which
can be used to represent knowledge of an application domain in a structured
and formal way. A fundamental characteristic of these languages is that they
have a formal semantics. Description logics are used for various
applications. Among them are: The representation of ontologies \cite{BaaderHS05}, natural
language processing \cite{Fehreretal94a}  and representation of the semantics of
UML class diagrams \cite{Berardi01reasoningon}.

We recall that description logics have as a common basis $ \mathcal {AL} $
enriched with different extensions: The description logic $ \mathcal {ALC} $,
subject of this work, adds  negation to $ \mathcal {AL} $. Other extensions add the
transitive closure of roles, number restrictions on roles, the notion of
sub-roles etc.. 
The formulas  $ C $ of $ \mathcal {ALC}$  logic, called \emph {concepts}, are constructed inductively by the following grammar:
\begin{alltt}
  \(C\) ::=   | \(A\)         (atomic concept)
           | \(\top\)         (universal concept  Top)
           | \(\bot\)         (empty concept  Bottom)
           | \(\neg C\)       (negation)
           | \(C \sqcap C\)    (conjunction)
           | \(C \sqcup C\)    (disjunction)
           | \(\forall r. C\)     (universal quantifier)
           | \(\exists r. C\)     (existential quantifier) \end{alltt}
Here, $ A \in NC $ is an atomic concept name, and $ r \in NR $  is a role  name. A role is a binary relation between instances of a concept. $ \forall $
and $ \exists $ are (multi-)modal operators, similar to $ \Box $ and
$\Diamond $ in traditional modal logics. 



%% file: Syntax.tex
\begin{isabellebody}%
\def\isabellecontext{Syntax}%
\isadelimtheory
\endisadelimtheory
\isatagtheory
\endisatagtheory
{\isafoldtheory}%
\isadelimtheory
\endisadelimtheory
\isamarkupsection{Syntax of  $\mathcal{ALC}$ \label{sec:syntax}%
}
\isamarkuptrue%
\begin{isamarkuptext}%
We now give details of the formal definition of the logic $ \mathcal {ALC} $. The type of roles, defined by:%
\end{isamarkuptext}%
\isamarkuptrue%
\ \isacommand{datatype}\isamarkupfalse%
\ {\isaliteral{27}{\isacharprime}}nr\ role\ {\isaliteral{3D}{\isacharequal}}\ AtomR\ {\isaliteral{27}{\isacharprime}}nr%
\begin{isamarkuptext}%
It only has a single constructor, but is easily expandable to accommodate more complex logic. Type definitions are parameterized by the type of role names \isa{{\isaliteral{27}{\isacharprime}}nr} and atomic concepts \isa{{\isaliteral{27}{\isacharprime}}nc}. Following the grammar of \secref {log_descr}, here is the type definition of $ \mathcal { ALC} $-concepts:%
\end{isamarkuptext}%
\isamarkuptrue%
\isacommand{datatype}\isamarkupfalse%
\ {\isaliteral{28}{\isacharparenleft}}{\isaliteral{27}{\isacharprime}}nr{\isaliteral{2C}{\isacharcomma}}\ {\isaliteral{27}{\isacharprime}}nc{\isaliteral{29}{\isacharparenright}}\ concept\ {\isaliteral{3D}{\isacharequal}}\isanewline
\ \ \ \ AtomC\ {\isaliteral{27}{\isacharprime}}nc\isanewline
\ \ {\isaliteral{7C}{\isacharbar}}\ Top\ \ \isanewline
\ \ {\isaliteral{7C}{\isacharbar}}\ Bottom\isanewline
\ \ {\isaliteral{7C}{\isacharbar}}\ NotC\ \ {\isaliteral{22}{\isachardoublequoteopen}}{\isaliteral{28}{\isacharparenleft}}{\isaliteral{28}{\isacharparenleft}}{\isaliteral{27}{\isacharprime}}nr{\isaliteral{2C}{\isacharcomma}}\ {\isaliteral{27}{\isacharprime}}nc{\isaliteral{29}{\isacharparenright}}\ concept{\isaliteral{29}{\isacharparenright}}{\isaliteral{22}{\isachardoublequoteclose}}\isanewline
\ \ {\isaliteral{7C}{\isacharbar}}\ AndC\ \ {\isaliteral{22}{\isachardoublequoteopen}}{\isaliteral{28}{\isacharparenleft}}{\isaliteral{28}{\isacharparenleft}}{\isaliteral{27}{\isacharprime}}nr{\isaliteral{2C}{\isacharcomma}}\ {\isaliteral{27}{\isacharprime}}nc{\isaliteral{29}{\isacharparenright}}\ concept{\isaliteral{29}{\isacharparenright}}{\isaliteral{22}{\isachardoublequoteclose}}\ {\isaliteral{22}{\isachardoublequoteopen}}{\isaliteral{28}{\isacharparenleft}}{\isaliteral{28}{\isacharparenleft}}{\isaliteral{27}{\isacharprime}}nr{\isaliteral{2C}{\isacharcomma}}\ {\isaliteral{27}{\isacharprime}}nc{\isaliteral{29}{\isacharparenright}}\ concept{\isaliteral{29}{\isacharparenright}}{\isaliteral{22}{\isachardoublequoteclose}}\isanewline
\ \ {\isaliteral{7C}{\isacharbar}}\ OrC\ \ {\isaliteral{22}{\isachardoublequoteopen}}{\isaliteral{28}{\isacharparenleft}}{\isaliteral{28}{\isacharparenleft}}{\isaliteral{27}{\isacharprime}}nr{\isaliteral{2C}{\isacharcomma}}\ {\isaliteral{27}{\isacharprime}}nc{\isaliteral{29}{\isacharparenright}}\ concept{\isaliteral{29}{\isacharparenright}}{\isaliteral{22}{\isachardoublequoteclose}}\ {\isaliteral{22}{\isachardoublequoteopen}}{\isaliteral{28}{\isacharparenleft}}{\isaliteral{28}{\isacharparenleft}}{\isaliteral{27}{\isacharprime}}nr{\isaliteral{2C}{\isacharcomma}}\ {\isaliteral{27}{\isacharprime}}nc{\isaliteral{29}{\isacharparenright}}\ concept{\isaliteral{29}{\isacharparenright}}{\isaliteral{22}{\isachardoublequoteclose}}\isanewline
\ \ {\isaliteral{7C}{\isacharbar}}\ AllC\ \ \ {\isaliteral{22}{\isachardoublequoteopen}}{\isaliteral{28}{\isacharparenleft}}{\isaliteral{27}{\isacharprime}}nr\ role{\isaliteral{29}{\isacharparenright}}{\isaliteral{22}{\isachardoublequoteclose}}\ {\isaliteral{22}{\isachardoublequoteopen}}{\isaliteral{28}{\isacharparenleft}}{\isaliteral{28}{\isacharparenleft}}{\isaliteral{27}{\isacharprime}}nr{\isaliteral{2C}{\isacharcomma}}\ {\isaliteral{27}{\isacharprime}}nc{\isaliteral{29}{\isacharparenright}}\ concept{\isaliteral{29}{\isacharparenright}}{\isaliteral{22}{\isachardoublequoteclose}}\isanewline
\ \ {\isaliteral{7C}{\isacharbar}}\ SomeC\ \ {\isaliteral{22}{\isachardoublequoteopen}}{\isaliteral{28}{\isacharparenleft}}{\isaliteral{27}{\isacharprime}}nr\ role{\isaliteral{29}{\isacharparenright}}{\isaliteral{22}{\isachardoublequoteclose}}\ {\isaliteral{22}{\isachardoublequoteopen}}{\isaliteral{28}{\isacharparenleft}}{\isaliteral{28}{\isacharparenleft}}{\isaliteral{27}{\isacharprime}}nr{\isaliteral{2C}{\isacharcomma}}\ {\isaliteral{27}{\isacharprime}}nc{\isaliteral{29}{\isacharparenright}}\ concept{\isaliteral{29}{\isacharparenright}}{\isaliteral{22}{\isachardoublequoteclose}}%
\isadelimtheory
\endisadelimtheory
\isatagtheory
\endisatagtheory
{\isafoldtheory}%
\isadelimtheory
\endisadelimtheory
\end{isabellebody}%

%% file: Semantic.tex
\begin{isabellebody}%
\def\isabellecontext{Semantic}%
\isadelimtheory
\endisadelimtheory
\isatagtheory
\endisatagtheory
{\isafoldtheory}%
\isadelimtheory
\endisadelimtheory
\isamarkupsection{Semantic of  $\mathcal{ALC}$ \label{sec:semantique}%
}
\isamarkuptrue%
\begin{isamarkuptext}%
Concepts are interpreted as subsets of a domain of interpretation $ \Delta_{\mathcal {I}} $ and roles as subsets of the product $ \Delta_{\mathcal {I}} \times \Delta_{\mathcal {I}} $. 
An interpretation $ \mathcal {I} $ is essentially a couple $ (\Delta_{\mathcal {I}}, .^ {\mathcal {I}}) $ where $ \Delta_{\mathcal {I}}$ is called the domain of interpretation and $.^{\mathcal {I}} $ is an interpretation function that maps an atomic concept  $A $ to subset $ A^ {\mathcal {I}} $ of $ \Delta_{\mathcal {I}} $ and a role  $ r $ to subset $ r ^{\mathcal {I}} $  of  $ \Delta_{\mathcal {I}} \times \Delta_{\mathcal {I}} $ . Its extension to other concept constructors is defined, in mathematical notation, as follows: \ \

\begin{tabular}[h]{ccl}
  $\top^{\mathcal{I}}$         & = & $\Delta_{\mathcal{I}}$\\
  $\bot^{\mathcal{I}}$         & = & $\emptyset $\\
  $(C \sqcap D)^{\mathcal{I}}$ & = & $C^{\mathcal{I}} \cap  D^{\mathcal{I}}$\\
  $(C \sqcup D)^{\mathcal{I}}$ & = & $C^{\mathcal{I}} \cup  D^{\mathcal{I}}$\\
  $(\neg C)^{\mathcal{I}}$     & = & $\Delta_{\mathcal{I}} - C^{\mathcal{I}}$\\
$(\forall r. C)^{\mathcal{I}}$& = & $\{x \in \Delta_{\mathcal{I}} / \forall y : (x,y) \in r^{\mathcal{I}} \rightarrow y \in C^{\mathcal{I}} \} $ \\
$(\exists r. C)^{\mathcal{I}}$ & = & $\{ x \in \Delta_{\mathcal{I}} / \exists y :
        (x,y) \in r^{\mathcal{I}} \wedge y \in C^{\mathcal{I}} \} $ \\ 
\end{tabular}%
\end{isamarkuptext}%
\isamarkuptrue%
\begin{isamarkuptext}%
The type \emph{domtype} is the type of  elements of the interpretation domain. Then The  interpretation is defined as follows:%
\end{isamarkuptext}%
\isamarkuptrue%
\isacommand{record}\isamarkupfalse%
\ {\isaliteral{28}{\isacharparenleft}}{\isaliteral{27}{\isacharprime}}ni{\isaliteral{2C}{\isacharcomma}}\ {\isaliteral{27}{\isacharprime}}nr{\isaliteral{2C}{\isacharcomma}}\ {\isaliteral{27}{\isacharprime}}nc{\isaliteral{29}{\isacharparenright}}\ Interp\ {\isaliteral{3D}{\isacharequal}}\isanewline
\ \ idomain\ {\isaliteral{3A}{\isacharcolon}}{\isaliteral{3A}{\isacharcolon}}\ \ {\isaliteral{22}{\isachardoublequoteopen}}domtype\ set{\isaliteral{22}{\isachardoublequoteclose}}\isanewline
\ \ interp{\isaliteral{5F}{\isacharunderscore}}c\ {\isaliteral{3A}{\isacharcolon}}{\isaliteral{3A}{\isacharcolon}}\ {\isaliteral{22}{\isachardoublequoteopen}}{\isaliteral{27}{\isacharprime}}nc\ {\isaliteral{5C3C52696768746172726F773E}{\isasymRightarrow}}\ domtype\ set\ {\isaliteral{22}{\isachardoublequoteclose}}\isanewline
\ \ interp{\isaliteral{5F}{\isacharunderscore}}r\ {\isaliteral{3A}{\isacharcolon}}{\isaliteral{3A}{\isacharcolon}}\ {\isaliteral{22}{\isachardoublequoteopen}}{\isaliteral{27}{\isacharprime}}nr\ {\isaliteral{5C3C52696768746172726F773E}{\isasymRightarrow}}\ \ {\isaliteral{28}{\isacharparenleft}}domtype\ \ {\isaliteral{2A}{\isacharasterisk}}\ domtype{\isaliteral{29}{\isacharparenright}}\ set{\isaliteral{22}{\isachardoublequoteclose}}\isanewline
\ \ interp{\isaliteral{5F}{\isacharunderscore}}i\ {\isaliteral{3A}{\isacharcolon}}{\isaliteral{3A}{\isacharcolon}}\ {\isaliteral{22}{\isachardoublequoteopen}}{\isaliteral{27}{\isacharprime}}ni\ {\isaliteral{5C3C52696768746172726F773E}{\isasymRightarrow}}\ domtype\ {\isaliteral{22}{\isachardoublequoteclose}}%
\begin{isamarkuptext}%
The interpretation of roles in Isabelle is given by:%
\end{isamarkuptext}%
\isamarkuptrue%
\isacommand{fun}\isamarkupfalse%
\ interpR\ \ {\isaliteral{3A}{\isacharcolon}}{\isaliteral{3A}{\isacharcolon}}\ {\isaliteral{22}{\isachardoublequoteopen}}{\isaliteral{28}{\isacharparenleft}}{\isaliteral{27}{\isacharprime}}ni{\isaliteral{2C}{\isacharcomma}}{\isaliteral{27}{\isacharprime}}nr{\isaliteral{2C}{\isacharcomma}}{\isaliteral{27}{\isacharprime}}nc{\isaliteral{29}{\isacharparenright}}\ Interp\ {\isaliteral{5C3C52696768746172726F773E}{\isasymRightarrow}}\ {\isaliteral{27}{\isacharprime}}nr\ role\ {\isaliteral{5C3C52696768746172726F773E}{\isasymRightarrow}}\ {\isaliteral{28}{\isacharparenleft}}domtype\ {\isaliteral{2A}{\isacharasterisk}}\ domtype{\isaliteral{29}{\isacharparenright}}\ set{\isaliteral{22}{\isachardoublequoteclose}}\ \isakeyword{where}\isanewline
\ \ {\isaliteral{22}{\isachardoublequoteopen}}interpR\ \ i\ {\isaliteral{28}{\isacharparenleft}}AtomR\ b{\isaliteral{29}{\isacharparenright}}\ {\isaliteral{3D}{\isacharequal}}\ {\isaliteral{28}{\isacharparenleft}}interp{\isaliteral{5F}{\isacharunderscore}}r\ i{\isaliteral{29}{\isacharparenright}}\ \ b{\isaliteral{22}{\isachardoublequoteclose}}%
\begin{isamarkuptext}%
The interpretation of concepts is described by the function:%
\end{isamarkuptext}%
\isamarkuptrue%
\isacommand{fun}\isamarkupfalse%
\ interpC\ {\isaliteral{3A}{\isacharcolon}}{\isaliteral{3A}{\isacharcolon}}\ {\isaliteral{22}{\isachardoublequoteopen}}{\isaliteral{28}{\isacharparenleft}}{\isaliteral{27}{\isacharprime}}ni{\isaliteral{2C}{\isacharcomma}}\ {\isaliteral{27}{\isacharprime}}nr{\isaliteral{2C}{\isacharcomma}}\ {\isaliteral{27}{\isacharprime}}nc{\isaliteral{29}{\isacharparenright}}\ Interp\ {\isaliteral{5C3C52696768746172726F773E}{\isasymRightarrow}}\ {\isaliteral{28}{\isacharparenleft}}{\isaliteral{27}{\isacharprime}}nr{\isaliteral{2C}{\isacharcomma}}\ {\isaliteral{27}{\isacharprime}}nc{\isaliteral{29}{\isacharparenright}}\ concept\ {\isaliteral{5C3C52696768746172726F773E}{\isasymRightarrow}}\ domtype\ set{\isaliteral{22}{\isachardoublequoteclose}}\ \ \isakeyword{where}\isanewline
\ \ \ \ {\isaliteral{22}{\isachardoublequoteopen}}interpC\ i\ Bottom\ {\isaliteral{3D}{\isacharequal}}\ {\isaliteral{7B}{\isacharbraceleft}}{\isaliteral{7D}{\isacharbraceright}}{\isaliteral{22}{\isachardoublequoteclose}}\isanewline
\ \ {\isaliteral{7C}{\isacharbar}}\ {\isaliteral{22}{\isachardoublequoteopen}}interpC\ i\ Top\ {\isaliteral{3D}{\isacharequal}}\ UNIV\ {\isaliteral{22}{\isachardoublequoteclose}}\isanewline
\ \ {\isaliteral{7C}{\isacharbar}}\ {\isaliteral{22}{\isachardoublequoteopen}}interpC\ i\ {\isaliteral{28}{\isacharparenleft}}AtomC\ a{\isaliteral{29}{\isacharparenright}}\ {\isaliteral{3D}{\isacharequal}}\ interp{\isaliteral{5F}{\isacharunderscore}}c\ i\ a{\isaliteral{22}{\isachardoublequoteclose}}\isanewline
\ \ {\isaliteral{7C}{\isacharbar}}\ {\isaliteral{22}{\isachardoublequoteopen}}interpC\ i\ {\isaliteral{28}{\isacharparenleft}}AndC\ c{\isadigit{1}}\ c{\isadigit{2}}{\isaliteral{29}{\isacharparenright}}\ {\isaliteral{3D}{\isacharequal}}\ {\isaliteral{28}{\isacharparenleft}}interpC\ i\ c{\isadigit{1}}{\isaliteral{29}{\isacharparenright}}\ {\isaliteral{5C3C696E7465723E}{\isasyminter}}\ \ {\isaliteral{28}{\isacharparenleft}}interpC\ \ i\ c{\isadigit{2}}{\isaliteral{29}{\isacharparenright}}{\isaliteral{22}{\isachardoublequoteclose}}\isanewline
\ \ {\isaliteral{7C}{\isacharbar}}\ {\isaliteral{22}{\isachardoublequoteopen}}interpC\ i\ {\isaliteral{28}{\isacharparenleft}}OrC\ c{\isadigit{1}}\ c{\isadigit{2}}{\isaliteral{29}{\isacharparenright}}\ {\isaliteral{3D}{\isacharequal}}\ {\isaliteral{28}{\isacharparenleft}}interpC\ i\ c{\isadigit{1}}{\isaliteral{29}{\isacharparenright}}\ {\isaliteral{5C3C756E696F6E3E}{\isasymunion}}\ {\isaliteral{28}{\isacharparenleft}}interpC\ \ i\ c{\isadigit{2}}{\isaliteral{29}{\isacharparenright}}{\isaliteral{22}{\isachardoublequoteclose}}\isanewline
\ \ {\isaliteral{7C}{\isacharbar}}\ {\isaliteral{22}{\isachardoublequoteopen}}interpC\ i\ {\isaliteral{28}{\isacharparenleft}}NotC\ c{\isaliteral{29}{\isacharparenright}}\ {\isaliteral{3D}{\isacharequal}}\ {\isaliteral{2D}{\isacharminus}}\ {\isaliteral{28}{\isacharparenleft}}interpC\ \ i\ c{\isaliteral{29}{\isacharparenright}}{\isaliteral{22}{\isachardoublequoteclose}}\isanewline
\ \ {\isaliteral{7C}{\isacharbar}}\ {\isaliteral{22}{\isachardoublequoteopen}}interpC\ i\ {\isaliteral{28}{\isacharparenleft}}AllC\ r\ c{\isaliteral{29}{\isacharparenright}}\ {\isaliteral{3D}{\isacharequal}}\ {\isaliteral{7B}{\isacharbraceleft}}x\ {\isaliteral{2E}{\isachardot}}\ {\isaliteral{5C3C666F72616C6C3E}{\isasymforall}}y{\isaliteral{2E}{\isachardot}}\ {\isaliteral{28}{\isacharparenleft}}{\isaliteral{28}{\isacharparenleft}}x{\isaliteral{2C}{\isacharcomma}}y{\isaliteral{29}{\isacharparenright}}\ {\isaliteral{5C3C696E3E}{\isasymin}}\ {\isaliteral{28}{\isacharparenleft}}interpR\ i\ r{\isaliteral{29}{\isacharparenright}}\ {\isaliteral{5C3C6C6F6E6772696768746172726F773E}{\isasymlongrightarrow}}\ y\ {\isaliteral{5C3C696E3E}{\isasymin}}\ {\isaliteral{28}{\isacharparenleft}}interpC\ i\ c{\isaliteral{29}{\isacharparenright}}{\isaliteral{29}{\isacharparenright}}\ {\isaliteral{7D}{\isacharbraceright}}{\isaliteral{22}{\isachardoublequoteclose}}\ \isanewline
\ \ {\isaliteral{7C}{\isacharbar}}\ {\isaliteral{22}{\isachardoublequoteopen}}interpC\ i\ {\isaliteral{28}{\isacharparenleft}}SomeC\ r\ c{\isaliteral{29}{\isacharparenright}}\ {\isaliteral{3D}{\isacharequal}}{\isaliteral{7B}{\isacharbraceleft}}\ x{\isaliteral{2E}{\isachardot}}\ {\isaliteral{5C3C6578697374733E}{\isasymexists}}y{\isaliteral{2E}{\isachardot}}\ {\isaliteral{28}{\isacharparenleft}}{\isaliteral{28}{\isacharparenleft}}x{\isaliteral{2C}{\isacharcomma}}y{\isaliteral{29}{\isacharparenright}}\ {\isaliteral{5C3C696E3E}{\isasymin}}{\isaliteral{28}{\isacharparenleft}}interpR\ i\ \ r\ {\isaliteral{29}{\isacharparenright}}\ {\isaliteral{5C3C616E643E}{\isasymand}}\ y\ {\isaliteral{5C3C696E3E}{\isasymin}}\ {\isaliteral{28}{\isacharparenleft}}interpC\ i\ c\ {\isaliteral{29}{\isacharparenright}}{\isaliteral{29}{\isacharparenright}}{\isaliteral{7D}{\isacharbraceright}}{\isaliteral{22}{\isachardoublequoteclose}}%
\isadelimproof
\endisadelimproof
\isatagproof
\endisatagproof
{\isafoldproof}%
\isadelimproof
\endisadelimproof
\isadelimproof
\endisadelimproof
\isatagproof
\endisatagproof
{\isafoldproof}%
\isadelimproof
\endisadelimproof
\isadelimproof
\endisadelimproof
\isatagproof
\endisatagproof
{\isafoldproof}%
\isadelimproof
\endisadelimproof
\isadelimproof
\endisadelimproof
\isatagproof
\endisatagproof
{\isafoldproof}%
\isadelimproof
\endisadelimproof
\isadelimproof
\endisadelimproof
\isatagproof
\endisatagproof
{\isafoldproof}%
\isadelimproof
\endisadelimproof
\isadelimproof
\endisadelimproof
\isatagproof
\endisatagproof
{\isafoldproof}%
\isadelimproof
\endisadelimproof
\isadelimproof
\endisadelimproof
\isatagproof
\endisatagproof
{\isafoldproof}%
\isadelimproof
\endisadelimproof
\isadelimproof
\endisadelimproof
\isatagproof
\endisatagproof
{\isafoldproof}%
\isadelimproof
\endisadelimproof
\isadelimproof
\endisadelimproof
\isatagproof
\endisatagproof
{\isafoldproof}%
\isadelimproof
\endisadelimproof
\begin{isamarkuptext}%
An interpretation $ \mathcal {I} $ is a model of the concept  $ C $ if $C ^{\mathcal {I}} \ne \emptyset $. As given by the following Isabelle definition:%
\end{isamarkuptext}%
\isamarkuptrue%
\isacommand{definition}\isamarkupfalse%
\ is{\isaliteral{5F}{\isacharunderscore}}model\ {\isaliteral{3A}{\isacharcolon}}{\isaliteral{3A}{\isacharcolon}}\ {\isaliteral{22}{\isachardoublequoteopen}}{\isaliteral{28}{\isacharparenleft}}{\isaliteral{27}{\isacharprime}}ni{\isaliteral{2C}{\isacharcomma}}\ {\isaliteral{27}{\isacharprime}}nr{\isaliteral{2C}{\isacharcomma}}\ {\isaliteral{27}{\isacharprime}}nc{\isaliteral{29}{\isacharparenright}}\ Interp\ {\isaliteral{5C3C52696768746172726F773E}{\isasymRightarrow}}\ {\isaliteral{28}{\isacharparenleft}}{\isaliteral{27}{\isacharprime}}nr{\isaliteral{2C}{\isacharcomma}}\ {\isaliteral{27}{\isacharprime}}nc{\isaliteral{29}{\isacharparenright}}\ concept\ \ {\isaliteral{5C3C52696768746172726F773E}{\isasymRightarrow}}\ bool{\isaliteral{22}{\isachardoublequoteclose}}\isanewline
\ \ \isakeyword{where}\ \ \ {\isaliteral{22}{\isachardoublequoteopen}}\ is{\isaliteral{5F}{\isacharunderscore}}model\ \ i\ c\ \ {\isaliteral{5C3C65717569763E}{\isasymequiv}}\ {\isaliteral{28}{\isacharparenleft}}interpC\ i\ c{\isaliteral{29}{\isacharparenright}}\ {\isaliteral{5C3C6E6F7465713E}{\isasymnoteq}}\ {\isaliteral{7B}{\isacharbraceleft}}{\isaliteral{7D}{\isacharbraceright}}{\isaliteral{22}{\isachardoublequoteclose}}%
\isadelimproof
\endisadelimproof
\isatagproof
\endisatagproof
{\isafoldproof}%
\isadelimproof
\endisadelimproof
\isadelimproof
\endisadelimproof
\isatagproof
\endisatagproof
{\isafoldproof}%
\isadelimproof
\endisadelimproof
\begin{isamarkuptext}%
A concept $C$ is satisfiable if there exists an interpretation $ \mathcal {I} $ such that $ \mathcal {I} $ is a model of $ C $. This definition is written in Isabelle:%
\end{isamarkuptext}%
\isamarkuptrue%
\isacommand{definition}\isamarkupfalse%
\ satisfiable{\isaliteral{5F}{\isacharunderscore}}def\ {\isaliteral{3A}{\isacharcolon}}{\isaliteral{3A}{\isacharcolon}}\ {\isaliteral{22}{\isachardoublequoteopen}}{\isaliteral{28}{\isacharparenleft}}{\isaliteral{27}{\isacharprime}}nr{\isaliteral{2C}{\isacharcomma}}\ {\isaliteral{27}{\isacharprime}}nc{\isaliteral{29}{\isacharparenright}}\ concept\ {\isaliteral{5C3C52696768746172726F773E}{\isasymRightarrow}}\ bool{\isaliteral{22}{\isachardoublequoteclose}}\isanewline
\ \ \isakeyword{where}\ {\isaliteral{22}{\isachardoublequoteopen}}satisfiable{\isaliteral{5F}{\isacharunderscore}}def\ c\ \ {\isaliteral{5C3C65717569763E}{\isasymequiv}}\ \ {\isaliteral{5C3C6578697374733E}{\isasymexists}}\ i{\isaliteral{2E}{\isachardot}}\ is{\isaliteral{5F}{\isacharunderscore}}model\ i\ c\ \ {\isaliteral{22}{\isachardoublequoteclose}}%
\isadelimproof
\endisadelimproof
\isatagproof
\endisatagproof
{\isafoldproof}%
\isadelimproof
\endisadelimproof
\isadelimproof
\endisadelimproof
\isatagproof
\endisatagproof
{\isafoldproof}%
\isadelimproof
\endisadelimproof
\isadelimproof
\endisadelimproof
\isatagproof
\endisatagproof
{\isafoldproof}%
\isadelimproof
\endisadelimproof
\isadelimproof
\endisadelimproof
\isatagproof
\endisatagproof
{\isafoldproof}%
\isadelimproof
\endisadelimproof
\isadelimproof
\endisadelimproof
\isatagproof
\endisatagproof
{\isafoldproof}%
\isadelimproof
\endisadelimproof
\isadelimproof
\endisadelimproof
\isatagproof
\endisatagproof
{\isafoldproof}%
\isadelimproof
\endisadelimproof
\isadelimproof
\endisadelimproof
\isatagproof
\endisatagproof
{\isafoldproof}%
\isadelimproof
\endisadelimproof
\isadelimproof
\endisadelimproof
\isatagproof
\endisatagproof
{\isafoldproof}%
\isadelimproof
\endisadelimproof
\isadelimtheory
\endisadelimtheory
\isatagtheory
\endisatagtheory
{\isafoldtheory}%
\isadelimtheory
\endisadelimtheory
\end{isabellebody}%

%% file: Tableau.tex
\begin{isabellebody}%
\def\isabellecontext{Tableau}%
\isadelimtheory
\endisadelimtheory
\isatagtheory
\endisatagtheory
{\isafoldtheory}%
\isadelimtheory
\endisadelimtheory
\isamarkupsection{Semantic tableau rules \label{sec:tableau}%
}
\isamarkuptrue%
\begin{isamarkuptext}%
Our rules handle \emph {Abox} (corresponding to a branch in a tableau), which are sets of facts. A fact may be of the form \isa{x{\isaliteral{3A}{\isacharcolon}}\ C} for an individual \isa{x} and concept \isa{C}, or  \isa{R\ x\ y}, for individuals \isa{x{\isaliteral{2C}{\isacharcomma}}\ y} and role \isa{R}. It can therefore be defined by:%
\end{isamarkuptext}%
\isamarkuptrue%
\isacommand{datatype}\isamarkupfalse%
\ \ {\isaliteral{28}{\isacharparenleft}}{\isaliteral{27}{\isacharprime}}ni{\isaliteral{2C}{\isacharcomma}}{\isaliteral{27}{\isacharprime}}nr{\isaliteral{2C}{\isacharcomma}}{\isaliteral{27}{\isacharprime}}nc{\isaliteral{29}{\isacharparenright}}\ fact\ \ {\isaliteral{3D}{\isacharequal}}\ \isanewline
\ \ \ \ Inst\ {\isaliteral{22}{\isachardoublequoteopen}}{\isaliteral{28}{\isacharparenleft}}{\isaliteral{27}{\isacharprime}}ni{\isaliteral{29}{\isacharparenright}}{\isaliteral{22}{\isachardoublequoteclose}}\ {\isaliteral{22}{\isachardoublequoteopen}}{\isaliteral{28}{\isacharparenleft}}{\isaliteral{28}{\isacharparenleft}}{\isaliteral{27}{\isacharprime}}nr{\isaliteral{2C}{\isacharcomma}}{\isaliteral{27}{\isacharprime}}nc{\isaliteral{29}{\isacharparenright}}\ concept{\isaliteral{29}{\isacharparenright}}{\isaliteral{22}{\isachardoublequoteclose}}\isanewline
\ \ {\isaliteral{7C}{\isacharbar}}\ Rel\ \ {\isaliteral{22}{\isachardoublequoteopen}}{\isaliteral{28}{\isacharparenleft}}{\isaliteral{27}{\isacharprime}}nr\ role{\isaliteral{29}{\isacharparenright}}{\isaliteral{22}{\isachardoublequoteclose}}\ {\isaliteral{22}{\isachardoublequoteopen}}{\isaliteral{28}{\isacharparenleft}}{\isaliteral{27}{\isacharprime}}ni{\isaliteral{29}{\isacharparenright}}{\isaliteral{22}{\isachardoublequoteclose}}\ {\isaliteral{22}{\isachardoublequoteopen}}{\isaliteral{28}{\isacharparenleft}}{\isaliteral{27}{\isacharprime}}ni{\isaliteral{29}{\isacharparenright}}{\isaliteral{22}{\isachardoublequoteclose}}\isanewline
\ \isanewline
\isacommand{type{\isaliteral{5F}{\isacharunderscore}}synonym}\isamarkupfalse%
{\isaliteral{28}{\isacharparenleft}}{\isaliteral{27}{\isacharprime}}ni{\isaliteral{2C}{\isacharcomma}}{\isaliteral{27}{\isacharprime}}nr{\isaliteral{2C}{\isacharcomma}}{\isaliteral{27}{\isacharprime}}nc{\isaliteral{29}{\isacharparenright}}\ abox\ {\isaliteral{3D}{\isacharequal}}\ {\isaliteral{22}{\isachardoublequoteopen}}{\isaliteral{28}{\isacharparenleft}}{\isaliteral{28}{\isacharparenleft}}{\isaliteral{27}{\isacharprime}}ni{\isaliteral{2C}{\isacharcomma}}{\isaliteral{27}{\isacharprime}}nr{\isaliteral{2C}{\isacharcomma}}{\isaliteral{27}{\isacharprime}}nc{\isaliteral{29}{\isacharparenright}}\ fact{\isaliteral{29}{\isacharparenright}}\ set{\isaliteral{22}{\isachardoublequoteclose}}%
\begin{isamarkuptext}%
It is now easy to define an interpretation that satisfies a fact and  an \emph{Abox}:%
\end{isamarkuptext}%
\isamarkuptrue%
\isacommand{fun}\isamarkupfalse%
\ satisfies{\isaliteral{5F}{\isacharunderscore}}fact\ {\isaliteral{3A}{\isacharcolon}}{\isaliteral{3A}{\isacharcolon}}\ {\isaliteral{22}{\isachardoublequoteopen}}{\isaliteral{28}{\isacharparenleft}}{\isaliteral{27}{\isacharprime}}ni{\isaliteral{2C}{\isacharcomma}}{\isaliteral{27}{\isacharprime}}nr{\isaliteral{2C}{\isacharcomma}}{\isaliteral{27}{\isacharprime}}nc{\isaliteral{29}{\isacharparenright}}\ Interp\ {\isaliteral{5C3C52696768746172726F773E}{\isasymRightarrow}}\ {\isaliteral{28}{\isacharparenleft}}{\isaliteral{27}{\isacharprime}}ni{\isaliteral{2C}{\isacharcomma}}{\isaliteral{27}{\isacharprime}}nr{\isaliteral{2C}{\isacharcomma}}{\isaliteral{27}{\isacharprime}}nc{\isaliteral{29}{\isacharparenright}}\ fact\ {\isaliteral{5C3C52696768746172726F773E}{\isasymRightarrow}}\ bool{\isaliteral{22}{\isachardoublequoteclose}}\isanewline
\ \isakeyword{where}\ \ {\isaliteral{22}{\isachardoublequoteopen}}satisfies{\isaliteral{5F}{\isacharunderscore}}fact\ icr\ {\isaliteral{28}{\isacharparenleft}}Inst\ x\ c{\isaliteral{29}{\isacharparenright}}\ {\isaliteral{3D}{\isacharequal}}\ {\isaliteral{28}{\isacharparenleft}}{\isaliteral{28}{\isacharparenleft}}interp{\isaliteral{5F}{\isacharunderscore}}i\ icr\ x{\isaliteral{29}{\isacharparenright}}\ {\isaliteral{5C3C696E3E}{\isasymin}}\ {\isaliteral{28}{\isacharparenleft}}interpC\ icr\ c{\isaliteral{29}{\isacharparenright}}{\isaliteral{29}{\isacharparenright}}{\isaliteral{22}{\isachardoublequoteclose}}\isanewline
\ \ \ \ \ \ \ {\isaliteral{7C}{\isacharbar}}{\isaliteral{22}{\isachardoublequoteopen}}satisfies{\isaliteral{5F}{\isacharunderscore}}fact\ icr\ {\isaliteral{28}{\isacharparenleft}}Rel\ r\ x\ y{\isaliteral{29}{\isacharparenright}}{\isaliteral{3D}{\isacharequal}}\ {\isaliteral{28}{\isacharparenleft}}{\isaliteral{28}{\isacharparenleft}}interp{\isaliteral{5F}{\isacharunderscore}}i\ icr\ x{\isaliteral{2C}{\isacharcomma}}\ interp{\isaliteral{5F}{\isacharunderscore}}i\ icr\ y{\isaliteral{29}{\isacharparenright}}\ {\isaliteral{5C3C696E3E}{\isasymin}}\ {\isaliteral{28}{\isacharparenleft}}interpR\ icr\ r{\isaliteral{29}{\isacharparenright}}{\isaliteral{29}{\isacharparenright}}{\isaliteral{22}{\isachardoublequoteclose}}\isanewline
\isanewline
\isacommand{definition}\isamarkupfalse%
\ satisfiable{\isaliteral{5F}{\isacharunderscore}}abox\ {\isaliteral{3A}{\isacharcolon}}{\isaliteral{3A}{\isacharcolon}}\ {\isaliteral{22}{\isachardoublequoteopen}}{\isaliteral{28}{\isacharparenleft}}{\isaliteral{28}{\isacharparenleft}}{\isaliteral{27}{\isacharprime}}ni{\isaliteral{2C}{\isacharcomma}}{\isaliteral{27}{\isacharprime}}nr{\isaliteral{2C}{\isacharcomma}}{\isaliteral{27}{\isacharprime}}nc{\isaliteral{29}{\isacharparenright}}\ abox{\isaliteral{29}{\isacharparenright}}\ {\isaliteral{5C3C52696768746172726F773E}{\isasymRightarrow}}\ bool{\isaliteral{22}{\isachardoublequoteclose}}\ \ \isanewline
\ \ \isakeyword{where}\ \ {\isaliteral{22}{\isachardoublequoteopen}}satisfiable{\isaliteral{5F}{\isacharunderscore}}abox\ Ab\ {\isaliteral{3D}{\isacharequal}}\ {\isaliteral{28}{\isacharparenleft}}{\isaliteral{5C3C6578697374733E}{\isasymexists}}\ i{\isaliteral{2E}{\isachardot}}\ {\isaliteral{28}{\isacharparenleft}}{\isaliteral{5C3C666F72616C6C3E}{\isasymforall}}\ f{\isaliteral{5C3C696E3E}{\isasymin}}\ Ab{\isaliteral{2E}{\isachardot}}\ satisfies{\isaliteral{5F}{\isacharunderscore}}fact\ i\ f{\isaliteral{29}{\isacharparenright}}{\isaliteral{29}{\isacharparenright}}{\isaliteral{22}{\isachardoublequoteclose}}%
\begin{isamarkuptext}%
We can now describe the rules of the decision procedure. A rule is a relationship between two \emph {Abox}, the \emph {Abox} before and after the application of the rule:%
\end{isamarkuptext}%
\isamarkuptrue%
\isacommand{type{\isaliteral{5F}{\isacharunderscore}}synonym}\isamarkupfalse%
\ {\isaliteral{28}{\isacharparenleft}}{\isaliteral{27}{\isacharprime}}ni{\isaliteral{2C}{\isacharcomma}}{\isaliteral{27}{\isacharprime}}nr{\isaliteral{2C}{\isacharcomma}}{\isaliteral{27}{\isacharprime}}nc{\isaliteral{29}{\isacharparenright}}rule\ {\isaliteral{3D}{\isacharequal}}\ {\isaliteral{22}{\isachardoublequoteopen}}{\isaliteral{28}{\isacharparenleft}}{\isaliteral{28}{\isacharparenleft}}{\isaliteral{27}{\isacharprime}}ni{\isaliteral{2C}{\isacharcomma}}{\isaliteral{27}{\isacharprime}}nr{\isaliteral{2C}{\isacharcomma}}{\isaliteral{27}{\isacharprime}}nc{\isaliteral{29}{\isacharparenright}}abox{\isaliteral{29}{\isacharparenright}}{\isaliteral{5C3C52696768746172726F773E}{\isasymRightarrow}}\ {\isaliteral{28}{\isacharparenleft}}{\isaliteral{28}{\isacharparenleft}}{\isaliteral{27}{\isacharprime}}ni{\isaliteral{2C}{\isacharcomma}}{\isaliteral{27}{\isacharprime}}nr{\isaliteral{2C}{\isacharcomma}}{\isaliteral{27}{\isacharprime}}nc{\isaliteral{29}{\isacharparenright}}abox{\isaliteral{29}{\isacharparenright}}{\isaliteral{5C3C52696768746172726F773E}{\isasymRightarrow}}\ bool{\isaliteral{22}{\isachardoublequoteclose}}%
\begin{isamarkuptext}%
This same format is applicable to simple rules, described later, and the composite rules. This homogeneous format is useful for writing and verifying tactics. For example, we show the rule for the constructor \isa{AndC}:%
\end{isamarkuptext}%
\isamarkuptrue%
\isacommand{inductive}\isamarkupfalse%
\ \ Andrule\ \ {\isaliteral{3A}{\isacharcolon}}{\isaliteral{3A}{\isacharcolon}}\ {\isaliteral{22}{\isachardoublequoteopen}}{\isaliteral{28}{\isacharparenleft}}{\isaliteral{27}{\isacharprime}}ni{\isaliteral{2C}{\isacharcomma}}{\isaliteral{27}{\isacharprime}}nr{\isaliteral{2C}{\isacharcomma}}{\isaliteral{27}{\isacharprime}}nc{\isaliteral{29}{\isacharparenright}}\ rule{\isaliteral{22}{\isachardoublequoteclose}}\ \isakeyword{where}\isanewline
\ \ mk{\isaliteral{5F}{\isacharunderscore}}andrule{\isaliteral{3A}{\isacharcolon}}\ {\isaliteral{22}{\isachardoublequoteopen}}{\isaliteral{5C3C6C6272616B6B3E}{\isasymlbrakk}}Inst\ x\ {\isaliteral{28}{\isacharparenleft}}AndC\ c{\isadigit{1}}\ c{\isadigit{2}}{\isaliteral{29}{\isacharparenright}}{\isaliteral{5C3C696E3E}{\isasymin}}\ b{\isadigit{1}}{\isaliteral{3B}{\isacharsemicolon}}\ {\isaliteral{5C3C6E6F743E}{\isasymnot}}\ {\isaliteral{28}{\isacharparenleft}}{\isaliteral{28}{\isacharparenleft}}Inst\ x\ c{\isadigit{1}}{\isaliteral{29}{\isacharparenright}}{\isaliteral{5C3C696E3E}{\isasymin}}\ b{\isadigit{1}}\ {\isaliteral{5C3C616E643E}{\isasymand}}\ {\isaliteral{28}{\isacharparenleft}}Inst\ x\ c{\isadigit{2}}{\isaliteral{29}{\isacharparenright}}\ {\isaliteral{5C3C696E3E}{\isasymin}}\ b{\isadigit{1}}{\isaliteral{29}{\isacharparenright}}{\isaliteral{3B}{\isacharsemicolon}}\isanewline
\ \ b{\isadigit{2}}\ {\isaliteral{3D}{\isacharequal}}\ {\isaliteral{7B}{\isacharbraceleft}}Inst\ x\ c{\isadigit{2}}{\isaliteral{7D}{\isacharbraceright}}\ {\isaliteral{5C3C756E696F6E3E}{\isasymunion}}\ \ {\isaliteral{7B}{\isacharbraceleft}}Inst\ x\ c{\isadigit{1}}{\isaliteral{7D}{\isacharbraceright}}\ {\isaliteral{5C3C756E696F6E3E}{\isasymunion}}\ b{\isadigit{1}}{\isaliteral{5C3C726272616B6B3E}{\isasymrbrakk}}\ {\isaliteral{5C3C4C6F6E6772696768746172726F773E}{\isasymLongrightarrow}}\ Andrule\ b{\isadigit{1}}\ b{\isadigit{2}}{\isaliteral{22}{\isachardoublequoteclose}}%
\begin{isamarkuptext}%
It expresses that an instance of the concept \isa{{\isaliteral{28}{\isacharparenleft}}AndC\ c{\isadigit{1}}\ c{\isadigit{2}}{\isaliteral{29}{\isacharparenright}}}] must be located in the \emph {Abox} before applying the rule (this is the condition of applicability), the concept has not been decomposed, and the application of the rule adds the sub-concepts \isa{c{\isadigit{1}}} and \isa{c{\isadigit{2}}}. Of course, this rule is highly non-deterministic, since it does not indicate which instance of a conjunction rule will be applied. Making the calculation more deterministic is one of the goals of the \secref{implantation}.%
\end{isamarkuptext}%
\isamarkuptrue%
\isadelimproof
\endisadelimproof
\isatagproof
\endisatagproof
{\isafoldproof}%
\isadelimproof
\endisadelimproof
\isadelimproof
\endisadelimproof
\isatagproof
\endisatagproof
{\isafoldproof}%
\isadelimproof
\endisadelimproof
\isadelimproof
\endisadelimproof
\isatagproof
\endisatagproof
{\isafoldproof}%
\isadelimproof
\endisadelimproof
\begin{isamarkuptext}%
For space reasons, we cannot present all the rules in detail. They are reproduced in \tabref{rules}. \input{ruletable}
The constructor \isa{SomeC} requires special attention. As indicated in \tabref{rules}, the application of rule requires the use of a new variable.What first comes to mind is to postulate the existence of this variable with an existential quantifier in the precondition of the rule. However, this non-deterministic existential choice would be impossible to implement by any specific generator function, or lead to a very complex notion of correspondence of abstract and implemented states.  We therefore parameterize the rule with the generator function \isa{gen}, which is also used in the implementation (see \secref {implantation}).%
\end{isamarkuptext}%
\isamarkuptrue%
\isacommand{inductive}\isamarkupfalse%
\ Somerule{\isaliteral{5F}{\isacharunderscore}}gen{\isaliteral{3A}{\isacharcolon}}{\isaliteral{3A}{\isacharcolon}}\ {\isaliteral{22}{\isachardoublequoteopen}}{\isaliteral{28}{\isacharparenleft}}{\isaliteral{28}{\isacharparenleft}}{\isaliteral{27}{\isacharprime}}ni{\isaliteral{2C}{\isacharcomma}}{\isaliteral{27}{\isacharprime}}nr{\isaliteral{2C}{\isacharcomma}}{\isaliteral{27}{\isacharprime}}nc{\isaliteral{29}{\isacharparenright}}abox{\isaliteral{5C3C52696768746172726F773E}{\isasymRightarrow}}{\isaliteral{27}{\isacharprime}}ni{\isaliteral{29}{\isacharparenright}}{\isaliteral{5C3C52696768746172726F773E}{\isasymRightarrow}}{\isaliteral{28}{\isacharparenleft}}{\isaliteral{27}{\isacharprime}}ni{\isaliteral{2C}{\isacharcomma}}{\isaliteral{27}{\isacharprime}}nr{\isaliteral{2C}{\isacharcomma}}{\isaliteral{27}{\isacharprime}}nc{\isaliteral{29}{\isacharparenright}}abox\isanewline
\ \ \ \ \ \ \ \ \ \ \ \ \ \ \ \ \ \ \ {\isaliteral{5C3C52696768746172726F773E}{\isasymRightarrow}}{\isaliteral{28}{\isacharparenleft}}{\isaliteral{27}{\isacharprime}}ni{\isaliteral{2C}{\isacharcomma}}{\isaliteral{27}{\isacharprime}}nr{\isaliteral{2C}{\isacharcomma}}{\isaliteral{27}{\isacharprime}}nc{\isaliteral{29}{\isacharparenright}}abox{\isaliteral{5C3C52696768746172726F773E}{\isasymRightarrow}}bool{\isaliteral{22}{\isachardoublequoteclose}}\ \isakeyword{where}\isanewline
\ \ \ mk{\isaliteral{5F}{\isacharunderscore}}Somerule{\isaliteral{5F}{\isacharunderscore}}gen{\isaliteral{3A}{\isacharcolon}}{\isaliteral{22}{\isachardoublequoteopen}}{\isaliteral{5C3C6C6272616B6B3E}{\isasymlbrakk}}{\isaliteral{28}{\isacharparenleft}}Inst\ x\ {\isaliteral{28}{\isacharparenleft}}SomeC\ r\ c{\isadigit{1}}{\isaliteral{29}{\isacharparenright}}{\isaliteral{29}{\isacharparenright}}{\isaliteral{5C3C696E3E}{\isasymin}}\ b{\isadigit{1}}{\isaliteral{3B}{\isacharsemicolon}}\ {\isaliteral{5C3C666F72616C6C3E}{\isasymforall}}\ y{\isaliteral{2E}{\isachardot}}\ {\isaliteral{5C3C6E6F743E}{\isasymnot}}{\isaliteral{28}{\isacharparenleft}}{\isaliteral{28}{\isacharparenleft}}Rel\ r\ x\ y{\isaliteral{29}{\isacharparenright}}{\isaliteral{5C3C696E3E}{\isasymin}}\ b{\isadigit{1}}\ {\isaliteral{5C3C616E643E}{\isasymand}}\ {\isaliteral{28}{\isacharparenleft}}Inst\ y\ c{\isadigit{1}}{\isaliteral{29}{\isacharparenright}}\ {\isaliteral{5C3C696E3E}{\isasymin}}\ b{\isadigit{1}}{\isaliteral{29}{\isacharparenright}}{\isaliteral{3B}{\isacharsemicolon}}z{\isaliteral{3D}{\isacharequal}}\ gen\ b{\isadigit{1}}{\isaliteral{3B}{\isacharsemicolon}}\ b{\isadigit{2}}{\isaliteral{3D}{\isacharequal}}\ insert\ {\isaliteral{28}{\isacharparenleft}}Rel\ r\ x\ z{\isaliteral{29}{\isacharparenright}}\ {\isaliteral{28}{\isacharparenleft}}insert\ {\isaliteral{28}{\isacharparenleft}}Inst\ z\ c{\isadigit{1}}{\isaliteral{29}{\isacharparenright}}\ b{\isadigit{1}}{\isaliteral{29}{\isacharparenright}}{\isaliteral{5C3C726272616B6B3E}{\isasymrbrakk}}\isanewline
\ \ \ \ \ \ \ \ \ \ \ \ \ \ \ \ \ \ \ \ \ \ \ \ \ \ \ \ \ \ \ \ \ \ \ \ \ \ \ \ {\isaliteral{5C3C4C6F6E6772696768746172726F773E}{\isasymLongrightarrow}}\ Somerule{\isaliteral{5F}{\isacharunderscore}}gen\ gen\ b{\isadigit{1}}\ b{\isadigit{2}}{\isaliteral{22}{\isachardoublequoteclose}}%
\begin{isamarkuptext}%
In summary, our rules are:%
\end{isamarkuptext}%
\isamarkuptrue%
\isacommand{definition}\isamarkupfalse%
\ list{\isaliteral{5F}{\isacharunderscore}}alc{\isaliteral{5F}{\isacharunderscore}}rules{\isaliteral{3A}{\isacharcolon}}{\isaliteral{3A}{\isacharcolon}}{\isaliteral{22}{\isachardoublequoteopen}}{\isaliteral{28}{\isacharparenleft}}{\isaliteral{28}{\isacharparenleft}}{\isaliteral{28}{\isacharparenleft}}{\isaliteral{27}{\isacharprime}}ni{\isaliteral{2C}{\isacharcomma}}{\isaliteral{27}{\isacharprime}}nr{\isaliteral{2C}{\isacharcomma}}{\isaliteral{27}{\isacharprime}}nc{\isaliteral{29}{\isacharparenright}}abox{\isaliteral{29}{\isacharparenright}}\ {\isaliteral{5C3C52696768746172726F773E}{\isasymRightarrow}}\ {\isaliteral{27}{\isacharprime}}ni{\isaliteral{29}{\isacharparenright}}\ {\isaliteral{5C3C52696768746172726F773E}{\isasymRightarrow}}\ {\isaliteral{28}{\isacharparenleft}}{\isaliteral{28}{\isacharparenleft}}{\isaliteral{27}{\isacharprime}}ni{\isaliteral{2C}{\isacharcomma}}{\isaliteral{27}{\isacharprime}}nr{\isaliteral{2C}{\isacharcomma}}{\isaliteral{27}{\isacharprime}}nc{\isaliteral{29}{\isacharparenright}}rule{\isaliteral{29}{\isacharparenright}}list{\isaliteral{22}{\isachardoublequoteclose}}\isanewline
\ \ \isakeyword{where}\ \ {\isaliteral{22}{\isachardoublequoteopen}}list{\isaliteral{5F}{\isacharunderscore}}alc{\isaliteral{5F}{\isacharunderscore}}rules\ gen\ {\isaliteral{3D}{\isacharequal}}\ \ {\isaliteral{5B}{\isacharbrackleft}}\ Andrule{\isaliteral{2C}{\isacharcomma}}\ Orrule{\isaliteral{2C}{\isacharcomma}}\ Allrule{\isaliteral{2C}{\isacharcomma}}\ Somerule{\isaliteral{5F}{\isacharunderscore}}gen\ gen\ {\isaliteral{5D}{\isacharbrackright}}{\isaliteral{22}{\isachardoublequoteclose}}%
\begin{isamarkuptext}%
From these elementary rules, we can construct composite rules by application of rule constructors, such as the following:%
\end{isamarkuptext}%
\isamarkuptrue%
\isacommand{fun}\isamarkupfalse%
\ \ disj{\isaliteral{5F}{\isacharunderscore}}rule\ {\isaliteral{3A}{\isacharcolon}}{\isaliteral{3A}{\isacharcolon}}\ {\isaliteral{22}{\isachardoublequoteopen}}{\isaliteral{28}{\isacharparenleft}}{\isaliteral{27}{\isacharprime}}ni{\isaliteral{2C}{\isacharcomma}}{\isaliteral{27}{\isacharprime}}nr{\isaliteral{2C}{\isacharcomma}}{\isaliteral{27}{\isacharprime}}nc{\isaliteral{29}{\isacharparenright}}\ rule\ {\isaliteral{5C3C52696768746172726F773E}{\isasymRightarrow}}\ {\isaliteral{28}{\isacharparenleft}}{\isaliteral{27}{\isacharprime}}ni{\isaliteral{2C}{\isacharcomma}}{\isaliteral{27}{\isacharprime}}nr{\isaliteral{2C}{\isacharcomma}}{\isaliteral{27}{\isacharprime}}nc{\isaliteral{29}{\isacharparenright}}\ rule\ {\isaliteral{5C3C52696768746172726F773E}{\isasymRightarrow}}\ {\isaliteral{28}{\isacharparenleft}}{\isaliteral{27}{\isacharprime}}ni{\isaliteral{2C}{\isacharcomma}}{\isaliteral{27}{\isacharprime}}nr{\isaliteral{2C}{\isacharcomma}}{\isaliteral{27}{\isacharprime}}nc{\isaliteral{29}{\isacharparenright}}\ rule{\isaliteral{22}{\isachardoublequoteclose}}\isanewline
\ \ \isakeyword{where}\ \ \ {\isaliteral{22}{\isachardoublequoteopen}}disj{\isaliteral{5F}{\isacharunderscore}}rule\ r{\isadigit{1}}\ r{\isadigit{2}}\ {\isaliteral{3D}{\isacharequal}}\ {\isaliteral{28}{\isacharparenleft}}{\isaliteral{5C3C6C616D6264613E}{\isasymlambda}}\ \ a\ b{\isaliteral{2E}{\isachardot}}\ \ r{\isadigit{1}}\ a\ b\ \ {\isaliteral{5C3C6F723E}{\isasymor}}\ \ r{\isadigit{2}}\ a\ b{\isaliteral{29}{\isacharparenright}}{\isaliteral{22}{\isachardoublequoteclose}}%
\begin{isamarkuptext}%
It allows to define by recursion the function \isa{disj{\isaliteral{5F}{\isacharunderscore}}rule{\isaliteral{5F}{\isacharunderscore}}list} that converts a list of rules in a rule. Finally, we define the rule%
\end{isamarkuptext}%
\isamarkuptrue%
\isacommand{definition}\isamarkupfalse%
\ alc{\isaliteral{5F}{\isacharunderscore}}rule\ {\isaliteral{3A}{\isacharcolon}}{\isaliteral{3A}{\isacharcolon}}\ {\isaliteral{22}{\isachardoublequoteopen}}{\isaliteral{28}{\isacharparenleft}}{\isaliteral{28}{\isacharparenleft}}{\isaliteral{28}{\isacharparenleft}}{\isaliteral{27}{\isacharprime}}ni{\isaliteral{2C}{\isacharcomma}}{\isaliteral{27}{\isacharprime}}nr{\isaliteral{2C}{\isacharcomma}}{\isaliteral{27}{\isacharprime}}nc{\isaliteral{29}{\isacharparenright}}\ abox{\isaliteral{29}{\isacharparenright}}\ {\isaliteral{5C3C52696768746172726F773E}{\isasymRightarrow}}\ {\isaliteral{27}{\isacharprime}}ni{\isaliteral{29}{\isacharparenright}}\ {\isaliteral{5C3C52696768746172726F773E}{\isasymRightarrow}}\ {\isaliteral{28}{\isacharparenleft}}{\isaliteral{27}{\isacharprime}}ni{\isaliteral{2C}{\isacharcomma}}{\isaliteral{27}{\isacharprime}}nr{\isaliteral{2C}{\isacharcomma}}{\isaliteral{27}{\isacharprime}}nc{\isaliteral{29}{\isacharparenright}}\ rule{\isaliteral{22}{\isachardoublequoteclose}}\isanewline
\ \ \isakeyword{where}\ \ \ {\isaliteral{22}{\isachardoublequoteopen}}alc{\isaliteral{5F}{\isacharunderscore}}rule\ gen\ {\isaliteral{3D}{\isacharequal}}\ disj{\isaliteral{5F}{\isacharunderscore}}rule{\isaliteral{5F}{\isacharunderscore}}list\ {\isaliteral{28}{\isacharparenleft}}list{\isaliteral{5F}{\isacharunderscore}}alc{\isaliteral{5F}{\isacharunderscore}}rules\ gen{\isaliteral{29}{\isacharparenright}}{\isaliteral{22}{\isachardoublequoteclose}}%
\isadelimproof
\endisadelimproof
\isatagproof
\endisatagproof
{\isafoldproof}%
\isadelimproof
\endisadelimproof
\isadelimproof
\endisadelimproof
\isatagproof
\endisatagproof
{\isafoldproof}%
\isadelimproof
\endisadelimproof
\isadelimproof
\endisadelimproof
\isatagproof
\endisatagproof
{\isafoldproof}%
\isadelimproof
\endisadelimproof
\isadelimproof
\endisadelimproof
\isatagproof
\endisatagproof
{\isafoldproof}%
\isadelimproof
\endisadelimproof
\isadelimproof
\endisadelimproof
\isatagproof
\endisatagproof
{\isafoldproof}%
\isadelimproof
\endisadelimproof
\isadelimproof
\endisadelimproof
\isatagproof
\endisatagproof
{\isafoldproof}%
\isadelimproof
\endisadelimproof
\isadelimproof
\endisadelimproof
\isatagproof
\endisatagproof
{\isafoldproof}%
\isadelimproof
\endisadelimproof
\isadelimproof
\endisadelimproof
\isatagproof
\endisatagproof
{\isafoldproof}%
\isadelimproof
\endisadelimproof
\isadelimproof
\endisadelimproof
\isatagproof
\endisatagproof
{\isafoldproof}%
\isadelimproof
\endisadelimproof
\isadelimproof
\endisadelimproof
\isatagproof
\endisatagproof
{\isafoldproof}%
\isadelimproof
\endisadelimproof
\isadelimproof
\endisadelimproof
\isatagproof
\endisatagproof
{\isafoldproof}%
\isadelimproof
\endisadelimproof
\isadelimproof
\endisadelimproof
\isatagproof
\endisatagproof
{\isafoldproof}%
\isadelimproof
\endisadelimproof
\isadelimproof
\endisadelimproof
\isatagproof
\endisatagproof
{\isafoldproof}%
\isadelimproof
\endisadelimproof
\isadelimproof
\endisadelimproof
\isatagproof
\endisatagproof
{\isafoldproof}%
\isadelimproof
\endisadelimproof
\isadelimproof
\endisadelimproof
\isatagproof
\endisatagproof
{\isafoldproof}%
\isadelimproof
\endisadelimproof
\isadelimproof
\endisadelimproof
\isatagproof
\endisatagproof
{\isafoldproof}%
\isadelimproof
\endisadelimproof
\isadelimproof
\endisadelimproof
\isatagproof
\endisatagproof
{\isafoldproof}%
\isadelimproof
\endisadelimproof
\isadelimproof
\endisadelimproof
\isatagproof
\endisatagproof
{\isafoldproof}%
\isadelimproof
\endisadelimproof
\isadelimproof
\endisadelimproof
\isatagproof
\endisatagproof
{\isafoldproof}%
\isadelimproof
\endisadelimproof
\isadelimproof
\endisadelimproof
\isatagproof
\endisatagproof
{\isafoldproof}%
\isadelimproof
\endisadelimproof
\isadelimproof
\endisadelimproof
\isatagproof
\endisatagproof
{\isafoldproof}%
\isadelimproof
\endisadelimproof
\isadelimtheory
\endisadelimtheory
\isatagtheory
\endisatagtheory
{\isafoldtheory}%
\isadelimtheory
\endisadelimtheory
\end{isabellebody}%

%% file: ruletable.tex
\small
\begin{table*} 
\begin{tabular}[ht]{|c|c|p{4cm}|p{4cm}|}
\hline
\textbf{Rule} & \textbf{Condition}  & \textbf{Negative Appl Cond} & \textbf{Action}  \\
\hline
  $\rightarrow_{\sqcap}$ &
  $x: C_1 \sqcap C_2  \in {\cal A}$ &
  $x: C_1$ and $ x: C_2$ are not both in $ {\cal A}$ &
  ${\cal A} := {\cal A}\cup \{ x: C_1, x: C_2 \}$ \\
\hline
  $ \rightarrow_{\sqcup}$ & 
  $x: C_1 \sqcup C_2 \in {\cal A}$ &
 neither $x: C_1$ nor $x : C_2$ in ${\cal A}$ &
  ${\cal A} := {\cal A}\cup\{ x: C_1\}$ or ${\cal A} := {\cal A}\cup\{x: C_2\}$\\
\hline
  $ \rightarrow_{\forall}$ &
  $x: \forall r C \in {\cal A}$ &
  $r\; x\; y  \in  {\cal A}$ but  $y:C \notin  {\cal A}$ &
  ${\cal A} := {\cal A}\cup\{y: C\}$\\
\hline
  $\rightarrow_{\exists}$ &
  $x: \exists r C \in {\cal A}$ &
  $\neg \exists  y$ such that $r\; x\; y$ and $y : C$ are both in ${\cal A}$&
  ${\cal A} := {\cal A}\cup\{z: C , r\; x\; z \}$ Where  $z$ is a new variable\\
\hline
\end{tabular}
\caption { The decomposition rules for the method of semantic tableaux for $\cal ALC$}
  \label{tab:rules}
\end{table*}
\normalsize


%% file: Soundness.tex
\begin{isabellebody}%
\def\isabellecontext{Soundness}%
\isadelimtheory
\endisadelimtheory
\isatagtheory
\endisatagtheory
{\isafoldtheory}%
\isadelimtheory
\endisadelimtheory
\isamarkupsection{Soundness\label{sec:correction}%
}
\isamarkuptrue%
\begin{isamarkuptext}%
The first central property  of a system of rules is  soundness. A rule is called sound, if its conclusion is satisfiable then its premise is satisfiable:%
\end{isamarkuptext}%
\isamarkuptrue%
\isacommand{definition}\isamarkupfalse%
\ \ sound\ {\isaliteral{3A}{\isacharcolon}}{\isaliteral{3A}{\isacharcolon}}\ {\isaliteral{22}{\isachardoublequoteopen}}{\isaliteral{28}{\isacharparenleft}}{\isaliteral{27}{\isacharprime}}ni{\isaliteral{2C}{\isacharcomma}}{\isaliteral{27}{\isacharprime}}nr{\isaliteral{2C}{\isacharcomma}}{\isaliteral{27}{\isacharprime}}nc{\isaliteral{29}{\isacharparenright}}\ rule\ {\isaliteral{5C3C52696768746172726F773E}{\isasymRightarrow}}\ bool{\isaliteral{22}{\isachardoublequoteclose}}\isanewline
\ \isakeyword{where}\ {\isaliteral{22}{\isachardoublequoteopen}}sound\ r{\isaliteral{3D}{\isacharequal}}{\isaliteral{3D}{\isacharequal}}\ {\isaliteral{5C3C666F72616C6C3E}{\isasymforall}}\ A{\isadigit{1}}\ A{\isadigit{2}}{\isaliteral{2E}{\isachardot}}\ r\ A{\isadigit{1}}\ A{\isadigit{2}}{\isaliteral{5C3C6C6F6E6772696768746172726F773E}{\isasymlongrightarrow}}\ satisfiable{\isaliteral{5F}{\isacharunderscore}}abox\ A{\isadigit{2}}{\isaliteral{5C3C6C6F6E6772696768746172726F773E}{\isasymlongrightarrow}}\ satisfiable{\isaliteral{5F}{\isacharunderscore}}abox\ A{\isadigit{1}}{\isaliteral{22}{\isachardoublequoteclose}}%
\isadelimproof
\endisadelimproof
\isatagproof
\endisatagproof
{\isafoldproof}%
\isadelimproof
\endisadelimproof
\begin{isamarkuptext}%
It is easy to show that the elementary rules preserve soundness, and the disjunction of rules:%
\end{isamarkuptext}%
\isamarkuptrue%
\isacommand{lemma}\isamarkupfalse%
\ disj{\isaliteral{5F}{\isacharunderscore}}rule{\isaliteral{5F}{\isacharunderscore}}sound\ {\isaliteral{5B}{\isacharbrackleft}}simp{\isaliteral{5D}{\isacharbrackright}}{\isaliteral{3A}{\isacharcolon}}\ {\isaliteral{22}{\isachardoublequoteopen}}sound\ r{\isadigit{1}}\ {\isaliteral{5C3C4C6F6E6772696768746172726F773E}{\isasymLongrightarrow}}\ sound\ r{\isadigit{2}}\ {\isaliteral{5C3C4C6F6E6772696768746172726F773E}{\isasymLongrightarrow}}\ sound\ {\isaliteral{28}{\isacharparenleft}}disj{\isaliteral{5F}{\isacharunderscore}}rule\ r{\isadigit{1}}\ r{\isadigit{2}}{\isaliteral{29}{\isacharparenright}}{\isaliteral{22}{\isachardoublequoteclose}}%
\isadelimproof
\endisadelimproof
\isatagproof
\endisatagproof
{\isafoldproof}%
\isadelimproof
\endisadelimproof
\isadelimproof
\endisadelimproof
\isatagproof
\endisatagproof
{\isafoldproof}%
\isadelimproof
\endisadelimproof
\isadelimproof
\endisadelimproof
\isatagproof
\endisatagproof
{\isafoldproof}%
\isadelimproof
\endisadelimproof
\isadelimproof
\endisadelimproof
\isatagproof
\endisatagproof
{\isafoldproof}%
\isadelimproof
\endisadelimproof
\isadelimproof
\endisadelimproof
\isatagproof
\endisatagproof
{\isafoldproof}%
\isadelimproof
\endisadelimproof
\begin{isamarkuptext}%
or the transitive closure:%
\end{isamarkuptext}%
\isamarkuptrue%
\isacommand{lemma}\isamarkupfalse%
\ tranclp{\isaliteral{5F}{\isacharunderscore}}rule{\isaliteral{5F}{\isacharunderscore}}sound\ {\isaliteral{5B}{\isacharbrackleft}}simp{\isaliteral{5D}{\isacharbrackright}}{\isaliteral{3A}{\isacharcolon}}\ {\isaliteral{22}{\isachardoublequoteopen}}sound\ r\ {\isaliteral{5C3C4C6F6E6772696768746172726F773E}{\isasymLongrightarrow}}\ sound\ {\isaliteral{28}{\isacharparenleft}}r{\isaliteral{5E}{\isacharcircum}}{\isaliteral{2B}{\isacharplus}}{\isaliteral{2B}{\isacharplus}}{\isaliteral{29}{\isacharparenright}}{\isaliteral{22}{\isachardoublequoteclose}}%
\isadelimproof
\endisadelimproof
\isatagproof
\endisatagproof
{\isafoldproof}%
\isadelimproof
\endisadelimproof
\begin{isamarkuptext}%
Also the proof of soundness of the various rules offers no surprises:%
\end{isamarkuptext}%
\isamarkuptrue%
\isadelimproof
\endisadelimproof
\isatagproof
\endisatagproof
{\isafoldproof}%
\isadelimproof
\endisadelimproof
\isadelimproof
\endisadelimproof
\isatagproof
\endisatagproof
{\isafoldproof}%
\isadelimproof
\endisadelimproof
\isadelimproof
\endisadelimproof
\isatagproof
\endisatagproof
{\isafoldproof}%
\isadelimproof
\endisadelimproof
\isadelimproof
\endisadelimproof
\isatagproof
\endisatagproof
{\isafoldproof}%
\isadelimproof
\endisadelimproof
\isadelimproof
\endisadelimproof
\isatagproof
\endisatagproof
{\isafoldproof}%
\isadelimproof
\endisadelimproof
\isadelimproof
\endisadelimproof
\isatagproof
\endisatagproof
{\isafoldproof}%
\isadelimproof
\endisadelimproof
\isadelimproof
\endisadelimproof
\isatagproof
\endisatagproof
{\isafoldproof}%
\isadelimproof
\endisadelimproof
\isadelimproof
\endisadelimproof
\isatagproof
\endisatagproof
{\isafoldproof}%
\isadelimproof
\endisadelimproof
\isacommand{lemma}\isamarkupfalse%
\ alcrule{\isaliteral{5F}{\isacharunderscore}}sound\ {\isaliteral{5B}{\isacharbrackleft}}simp{\isaliteral{5D}{\isacharbrackright}}{\isaliteral{3A}{\isacharcolon}}\ {\isaliteral{22}{\isachardoublequoteopen}}sound\ {\isaliteral{28}{\isacharparenleft}}alc{\isaliteral{5F}{\isacharunderscore}}rule\ gen{\isaliteral{29}{\isacharparenright}}{\isaliteral{22}{\isachardoublequoteclose}}%
\isadelimproof
\endisadelimproof
\isatagproof
\endisatagproof
{\isafoldproof}%
\isadelimproof
\endisadelimproof
\isadelimtheory
\endisadelimtheory
\isatagtheory
\endisatagtheory
{\isafoldtheory}%
\isadelimtheory
\endisadelimtheory
\end{isabellebody}%

%% file: Completeness.tex
\begin{isabellebody}%
\def\isabellecontext{Completeness}%
\isadelimtheory
\endisadelimtheory
\isatagtheory
\endisatagtheory
{\isafoldtheory}%
\isadelimtheory
\endisadelimtheory
\isamarkupsection{Completeness  \label{sec:completude}%
}
\isamarkuptrue%
\begin{isamarkuptext}%
To prove completeness, we first define the notion of complete rule. A rule is complete if the satisfiability of \emph{Abox} \isa{A{\isadigit{1}}} implies that there exists at least one satisfiable \emph {Abox} \isa{A{\isadigit{2}}} obtained by rule application from \isa{A{\isadigit{1}}}.%
\end{isamarkuptext}%
\isamarkuptrue%
\isacommand{definition}\isamarkupfalse%
\ complete\ {\isaliteral{3A}{\isacharcolon}}{\isaliteral{3A}{\isacharcolon}}{\isaliteral{22}{\isachardoublequoteopen}}{\isaliteral{28}{\isacharparenleft}}{\isaliteral{27}{\isacharprime}}ni{\isaliteral{2C}{\isacharcomma}}{\isaliteral{27}{\isacharprime}}nr{\isaliteral{2C}{\isacharcomma}}{\isaliteral{27}{\isacharprime}}nc{\isaliteral{29}{\isacharparenright}}\ rule\ {\isaliteral{5C3C52696768746172726F773E}{\isasymRightarrow}}\ bool{\isaliteral{22}{\isachardoublequoteclose}}\ \isanewline
\ \ \ \isakeyword{where}\ {\isaliteral{22}{\isachardoublequoteopen}}complete\ r\ {\isaliteral{3D}{\isacharequal}}{\isaliteral{3D}{\isacharequal}}\ {\isaliteral{5C3C666F72616C6C3E}{\isasymforall}}\ A{\isadigit{1}}{\isaliteral{2E}{\isachardot}}{\isaliteral{5C3C6578697374733E}{\isasymexists}}\ A{\isadigit{2}}{\isaliteral{2E}{\isachardot}}\ satisfiable{\isaliteral{5F}{\isacharunderscore}}abox\ A{\isadigit{1}}\ {\isaliteral{5C3C6C6F6E6772696768746172726F773E}{\isasymlongrightarrow}}\ {\isaliteral{28}{\isacharparenleft}}\ r\ A{\isadigit{1}}\ A{\isadigit{2}}{\isaliteral{29}{\isacharparenright}}\isanewline
\ \ \ \ \ \ \ \ \ \ \ \ \ \ \ \ \ \ \ \ \ \ {\isaliteral{5C3C6C6F6E6772696768746172726F773E}{\isasymlongrightarrow}}\ satisfiable{\isaliteral{5F}{\isacharunderscore}}abox\ A{\isadigit{2}}{\isaliteral{22}{\isachardoublequoteclose}}%
\begin{isamarkuptext}%
We can show this property for each rule. For the rule $\rightarrow_{ \sqcap}$, we obtain:%
\end{isamarkuptext}%
\isamarkuptrue%
\isacommand{lemma}\isamarkupfalse%
\ and{\isaliteral{5F}{\isacharunderscore}}complete\ {\isaliteral{5B}{\isacharbrackleft}}simp{\isaliteral{5D}{\isacharbrackright}}{\isaliteral{3A}{\isacharcolon}}\ {\isaliteral{22}{\isachardoublequoteopen}}complete\ Andrule{\isaliteral{22}{\isachardoublequoteclose}}\ %
\isadelimproof
\endisadelimproof
\isatagproof
\endisatagproof
{\isafoldproof}%
\isadelimproof
\endisadelimproof
\isadelimproof
\endisadelimproof
\isatagproof
\endisatagproof
{\isafoldproof}%
\isadelimproof
\endisadelimproof
\isadelimproof
\endisadelimproof
\isatagproof
\endisatagproof
{\isafoldproof}%
\isadelimproof
\endisadelimproof
\isadelimproof
\endisadelimproof
\isatagproof
\endisatagproof
{\isafoldproof}%
\isadelimproof
\endisadelimproof
\isadelimproof
\endisadelimproof
\isatagproof
\endisatagproof
{\isafoldproof}%
\isadelimproof
\endisadelimproof
\isadelimproof
\endisadelimproof
\isatagproof
\endisatagproof
{\isafoldproof}%
\isadelimproof
\endisadelimproof
\isadelimproof
\endisadelimproof
\isatagproof
\endisatagproof
{\isafoldproof}%
\isadelimproof
\endisadelimproof
\isadelimproof
\endisadelimproof
\isatagproof
\endisatagproof
{\isafoldproof}%
\isadelimproof
\endisadelimproof
\isadelimproof
\endisadelimproof
\isatagproof
\endisatagproof
{\isafoldproof}%
\isadelimproof
\endisadelimproof
\isadelimproof
\endisadelimproof
\isatagproof
\endisatagproof
{\isafoldproof}%
\isadelimproof
\endisadelimproof
\isadelimproof
\endisadelimproof
\isatagproof
\endisatagproof
{\isafoldproof}%
\isadelimproof
\endisadelimproof
\isadelimproof
\endisadelimproof
\isatagproof
\endisatagproof
{\isafoldproof}%
\isadelimproof
\endisadelimproof
\isadelimproof
\endisadelimproof
\isatagproof
\endisatagproof
{\isafoldproof}%
\isadelimproof
\endisadelimproof
\begin{isamarkuptext}%
An \emph {Abox} is \emph{contradictory} if it contains a contradiction (clash), i.e, $x: C$ and  $x : \neg C$ or $x :\bot$.%
\end{isamarkuptext}%
\isamarkuptrue%
\isacommand{fun}\isamarkupfalse%
\ contains{\isaliteral{5F}{\isacharunderscore}}clash\ {\isaliteral{3A}{\isacharcolon}}{\isaliteral{3A}{\isacharcolon}}\ {\isaliteral{22}{\isachardoublequoteopen}}{\isaliteral{28}{\isacharparenleft}}{\isaliteral{27}{\isacharprime}}ni{\isaliteral{2C}{\isacharcomma}}{\isaliteral{27}{\isacharprime}}nr{\isaliteral{2C}{\isacharcomma}}{\isaliteral{27}{\isacharprime}}nc{\isaliteral{29}{\isacharparenright}}\ abox\ {\isaliteral{5C3C52696768746172726F773E}{\isasymRightarrow}}\ bool{\isaliteral{22}{\isachardoublequoteclose}}\isanewline
\ \ \ \ \isakeyword{where}\ \ {\isaliteral{22}{\isachardoublequoteopen}}contains{\isaliteral{5F}{\isacharunderscore}}clash\ AB\ {\isaliteral{3D}{\isacharequal}}\ \isanewline
\ \ \ \ \ {\isaliteral{28}{\isacharparenleft}}{\isaliteral{5C3C6578697374733E}{\isasymexists}}\ x\ c{\isaliteral{2E}{\isachardot}}\ {\isaliteral{28}{\isacharparenleft}}{\isaliteral{28}{\isacharparenleft}}Inst\ x\ c{\isaliteral{29}{\isacharparenright}}\ {\isaliteral{5C3C696E3E}{\isasymin}}\ AB\ {\isaliteral{5C3C616E643E}{\isasymand}}\ {\isaliteral{28}{\isacharparenleft}}Inst\ x\ {\isaliteral{28}{\isacharparenleft}}NotC\ c{\isaliteral{29}{\isacharparenright}}{\isaliteral{29}{\isacharparenright}}\ {\isaliteral{5C3C696E3E}{\isasymin}}\ AB{\isaliteral{29}{\isacharparenright}}\ {\isaliteral{5C3C6F723E}{\isasymor}}\ {\isaliteral{28}{\isacharparenleft}}{\isaliteral{28}{\isacharparenleft}}Inst\ x\ Bottom{\isaliteral{29}{\isacharparenright}}\ {\isaliteral{5C3C696E3E}{\isasymin}}\ AB{\isaliteral{29}{\isacharparenright}}{\isaliteral{29}{\isacharparenright}}{\isaliteral{22}{\isachardoublequoteclose}}\ %
\begin{isamarkuptext}%
The fundamental property of the correctness of the tableau algorithm  is that if the \emph{Abox} is closed (contains a clash) then it is unsatisfiable:%
\end{isamarkuptext}%
\isamarkuptrue%
\isacommand{lemma}\isamarkupfalse%
\ content{\isaliteral{5F}{\isacharunderscore}}clash{\isaliteral{5F}{\isacharunderscore}}not{\isaliteral{5F}{\isacharunderscore}}satisfiable{\isaliteral{3A}{\isacharcolon}}{\isaliteral{22}{\isachardoublequoteopen}}{\isaliteral{5C3C6C6272616B6B3E}{\isasymlbrakk}}contains{\isaliteral{5F}{\isacharunderscore}}clash\ AB{\isaliteral{3B}{\isacharsemicolon}}satisfiable{\isaliteral{5F}{\isacharunderscore}}abox\ AB{\isaliteral{5C3C726272616B6B3E}{\isasymrbrakk}}\ {\isaliteral{5C3C4C6F6E6772696768746172726F773E}{\isasymLongrightarrow}}\ False{\isaliteral{22}{\isachardoublequoteclose}}\ %
\isadelimproof
\endisadelimproof
\isatagproof
\endisatagproof
{\isafoldproof}%
\isadelimproof
\endisadelimproof
\isadelimproof
\endisadelimproof
\isatagproof
\endisatagproof
{\isafoldproof}%
\isadelimproof
\endisadelimproof
\begin{isamarkuptext}%
An \emph{Abox} is saturated for a rule if the rule is not applicable to it.%
\end{isamarkuptext}%
\isamarkuptrue%
\isacommand{definition}\isamarkupfalse%
\ saturated\ {\isaliteral{3A}{\isacharcolon}}{\isaliteral{3A}{\isacharcolon}}\ {\isaliteral{22}{\isachardoublequoteopen}}{\isaliteral{28}{\isacharparenleft}}{\isaliteral{27}{\isacharprime}}ni{\isaliteral{2C}{\isacharcomma}}{\isaliteral{27}{\isacharprime}}nr{\isaliteral{2C}{\isacharcomma}}{\isaliteral{27}{\isacharprime}}nc{\isaliteral{29}{\isacharparenright}}\ abox\ {\isaliteral{5C3C52696768746172726F773E}{\isasymRightarrow}}\ {\isaliteral{28}{\isacharparenleft}}{\isaliteral{27}{\isacharprime}}ni{\isaliteral{2C}{\isacharcomma}}{\isaliteral{27}{\isacharprime}}nr{\isaliteral{2C}{\isacharcomma}}{\isaliteral{27}{\isacharprime}}nc{\isaliteral{29}{\isacharparenright}}\ rule\ {\isaliteral{5C3C52696768746172726F773E}{\isasymRightarrow}}\ bool{\isaliteral{22}{\isachardoublequoteclose}}\isanewline
\ \ \ \ \isakeyword{where}\ {\isaliteral{22}{\isachardoublequoteopen}}saturated\ AB{\isadigit{1}}\ r\ {\isaliteral{5C3C65717569763E}{\isasymequiv}}\ {\isaliteral{28}{\isacharparenleft}}{\isaliteral{5C3C666F72616C6C3E}{\isasymforall}}\ AB{\isadigit{2}}{\isaliteral{2E}{\isachardot}}\ {\isaliteral{5C3C6E6F743E}{\isasymnot}}\ {\isaliteral{28}{\isacharparenleft}}r\ AB{\isadigit{1}}\ AB{\isadigit{2}}{\isaliteral{29}{\isacharparenright}}{\isaliteral{29}{\isacharparenright}}{\isaliteral{22}{\isachardoublequoteclose}}%
\begin{isamarkuptext}%
Finally, if an \emph{Abox} $A$ is saturated and not contradictory, then it is satisfiable. In this case, there is an interpretation that satisfies $A$, which is called the \emph{canonical interpretation} $ \mathcal{I}_A $, whose components are defined as follows:
\begin{enumerate}
\item The interpretation domain $ \Delta_{\mathcal {I}_A} $ is the set of all individuals included in $ A $ 
\item For each concept name $ P $ we define
  $P _ {\mathcal{I}_A} = \{ x | (x: P) \in A \}$
\item For each role name $ r $ we define  $r _ {\mathcal{I}_A} = \{ (x, y) | r(x,y) \in A \}$
\end{enumerate}%
\end{isamarkuptext}%
\isamarkuptrue%
\isadelimproof
\endisadelimproof
\isatagproof
\endisatagproof
{\isafoldproof}%
\isadelimproof
\endisadelimproof
\isadelimproof
\endisadelimproof
\isatagproof
\endisatagproof
{\isafoldproof}%
\isadelimproof
\endisadelimproof
\isadelimproof
\endisadelimproof
\isatagproof
\endisatagproof
{\isafoldproof}%
\isadelimproof
\endisadelimproof
\isadelimproof
\endisadelimproof
\isatagproof
\endisatagproof
{\isafoldproof}%
\isadelimproof
\endisadelimproof
\isadelimproof
\endisadelimproof
\isatagproof
\endisatagproof
{\isafoldproof}%
\isadelimproof
\endisadelimproof
\isadelimproof
\endisadelimproof
\isatagproof
\endisatagproof
{\isafoldproof}%
\isadelimproof
\endisadelimproof
\isadelimproof
\endisadelimproof
\isatagproof
\endisatagproof
{\isafoldproof}%
\isadelimproof
\endisadelimproof
\isadelimproof
\endisadelimproof
\isatagproof
\endisatagproof
{\isafoldproof}%
\isadelimproof
\endisadelimproof
\isadelimproof
\endisadelimproof
\isatagproof
\endisatagproof
{\isafoldproof}%
\isadelimproof
\endisadelimproof
\isadelimproof
\endisadelimproof
\isatagproof
\endisatagproof
{\isafoldproof}%
\isadelimproof
\endisadelimproof
\isadelimproof
\endisadelimproof
\isatagproof
\endisatagproof
{\isafoldproof}%
\isadelimproof
\endisadelimproof
\isadelimproof
\endisadelimproof
\isatagproof
\endisatagproof
{\isafoldproof}%
\isadelimproof
\endisadelimproof
\isadelimproof
\endisadelimproof
\isatagproof
\endisatagproof
{\isafoldproof}%
\isadelimproof
\endisadelimproof
\isadelimproof
\endisadelimproof
\isatagproof
\endisatagproof
{\isafoldproof}%
\isadelimproof
\endisadelimproof
\isadelimproof
\endisadelimproof
\isatagproof
\endisatagproof
{\isafoldproof}%
\isadelimproof
\endisadelimproof
\isadelimproof
\endisadelimproof
\isatagproof
\endisatagproof
{\isafoldproof}%
\isadelimproof
\endisadelimproof
\isadelimproof
\endisadelimproof
\isatagproof
\endisatagproof
{\isafoldproof}%
\isadelimproof
\endisadelimproof
\isadelimproof
\endisadelimproof
\isatagproof
\endisatagproof
{\isafoldproof}%
\isadelimproof
\endisadelimproof
\isadelimproof
\endisadelimproof
\isatagproof
\endisatagproof
{\isafoldproof}%
\isadelimproof
\endisadelimproof
\isadelimproof
\endisadelimproof
\isatagproof
\endisatagproof
{\isafoldproof}%
\isadelimproof
\endisadelimproof
\begin{isamarkuptext}%
Now the goal is to prove that if an \emph {Abox} is not contradictory and saturated, then it is satisfiable by the canonical interpretation.%
\end{isamarkuptext}%
\isamarkuptrue%
\isacommand{lemma}\isamarkupfalse%
\ canon{\isaliteral{5F}{\isacharunderscore}}interp{\isaliteral{5F}{\isacharunderscore}}sat{\isaliteral{5F}{\isacharunderscore}}fact{\isaliteral{3A}{\isacharcolon}}{\isaliteral{22}{\isachardoublequoteopen}}{\isaliteral{5C3C6C6272616B6B3E}{\isasymlbrakk}}inj\ i{\isaliteral{3B}{\isacharsemicolon}}\ {\isaliteral{5C3C6E6F743E}{\isasymnot}}contains{\isaliteral{5F}{\isacharunderscore}}clash\ AB{\isaliteral{3B}{\isacharsemicolon}}\ saturated\ AB\ {\isaliteral{28}{\isacharparenleft}}alc{\isaliteral{5F}{\isacharunderscore}}rule\ gen{\isaliteral{29}{\isacharparenright}}{\isaliteral{3B}{\isacharsemicolon}}\isanewline
\ \ \ \ \ \ \ \ \ \ \ \ \ \ \ \ \ \ \ \ \ \ \ is{\isaliteral{5F}{\isacharunderscore}}Normal{\isaliteral{5F}{\isacharunderscore}}Abox\ AB{\isaliteral{3B}{\isacharsemicolon}}\ f{\isaliteral{5C3C696E3E}{\isasymin}}\ AB{\isaliteral{5C3C726272616B6B3E}{\isasymrbrakk}}\ {\isaliteral{5C3C4C6F6E6772696768746172726F773E}{\isasymLongrightarrow}}\ satisfies{\isaliteral{5F}{\isacharunderscore}}fact\ {\isaliteral{28}{\isacharparenleft}}canon{\isaliteral{5F}{\isacharunderscore}}interp\ i\ AB{\isaliteral{29}{\isacharparenright}}\ f{\isaliteral{22}{\isachardoublequoteclose}}%
\isadelimproof
\endisadelimproof
\isatagproof
\endisatagproof
{\isafoldproof}%
\isadelimproof
\endisadelimproof
\isadelimtheory
\endisadelimtheory
\isatagtheory
\endisatagtheory
{\isafoldtheory}%
\isadelimtheory
\endisadelimtheory
\end{isabellebody}%

%% file: Implementation.tex
\begin{isabellebody}%
\def\isabellecontext{Implementation}%
\isadelimtheory
\endisadelimtheory
\isatagtheory
\endisatagtheory
{\isafoldtheory}%
\isadelimtheory
\endisadelimtheory
\isamarkupsection{Implementation \label{sec:implantation}%
}
\isamarkuptrue%
\begin{isamarkuptext}%
In this section, we propose an implementation, i.e. an executable proof procedure for the description logic $ \mathcal {ALC} $. It is based on lists as data structure to implement the \emph {Abox}. The tableau is encoded as a list of \emph {Abox}. The different rules are defined as functions whose argument is list \isa{abox{\isaliteral{5F}{\isacharunderscore}}impl} and return a list of \emph {Abox} (Tableau).%
\end{isamarkuptext}%
\isamarkuptrue%
\begin{isamarkuptext}%
The \emph {Abox} is implemented as a list of facts (\emph {abox\_impl}).%
\end{isamarkuptext}%
\isamarkuptrue%
\isacommand{type{\isaliteral{5F}{\isacharunderscore}}synonym}\isamarkupfalse%
\ {\isaliteral{28}{\isacharparenleft}}{\isaliteral{27}{\isacharprime}}ni{\isaliteral{2C}{\isacharcomma}}{\isaliteral{27}{\isacharprime}}nr{\isaliteral{2C}{\isacharcomma}}{\isaliteral{27}{\isacharprime}}nc{\isaliteral{29}{\isacharparenright}}\ abox{\isaliteral{5F}{\isacharunderscore}}impl\ {\isaliteral{3D}{\isacharequal}}\ \ {\isaliteral{22}{\isachardoublequoteopen}}{\isaliteral{28}{\isacharparenleft}}{\isaliteral{28}{\isacharparenleft}}{\isaliteral{27}{\isacharprime}}ni{\isaliteral{2C}{\isacharcomma}}{\isaliteral{27}{\isacharprime}}nr{\isaliteral{2C}{\isacharcomma}}{\isaliteral{27}{\isacharprime}}nc{\isaliteral{29}{\isacharparenright}}\ fact{\isaliteral{29}{\isacharparenright}}\ list{\isaliteral{22}{\isachardoublequoteclose}}%
\begin{isamarkuptext}%
The implementation of a rule is defined as a function that transforms an \emph {Abox\_impl} to a list of \emph {Abox\_impl}, just called a tableau.%
\end{isamarkuptext}%
\isamarkuptrue%
\isacommand{type{\isaliteral{5F}{\isacharunderscore}}synonym}\isamarkupfalse%
\ {\isaliteral{28}{\isacharparenleft}}{\isaliteral{27}{\isacharprime}}ni{\isaliteral{2C}{\isacharcomma}}{\isaliteral{27}{\isacharprime}}nr{\isaliteral{2C}{\isacharcomma}}{\isaliteral{27}{\isacharprime}}nc{\isaliteral{29}{\isacharparenright}}rule{\isaliteral{5F}{\isacharunderscore}}impl{\isaliteral{3D}{\isacharequal}}\isanewline
\ \ \ \ \ \ \ {\isaliteral{22}{\isachardoublequoteopen}}{\isaliteral{28}{\isacharparenleft}}{\isaliteral{27}{\isacharprime}}ni{\isaliteral{2C}{\isacharcomma}}{\isaliteral{27}{\isacharprime}}nr{\isaliteral{2C}{\isacharcomma}}{\isaliteral{27}{\isacharprime}}nc{\isaliteral{29}{\isacharparenright}}abox{\isaliteral{5F}{\isacharunderscore}}impl{\isaliteral{5C3C52696768746172726F773E}{\isasymRightarrow}}{\isaliteral{28}{\isacharparenleft}}{\isaliteral{27}{\isacharprime}}ni{\isaliteral{2C}{\isacharcomma}}{\isaliteral{27}{\isacharprime}}nr{\isaliteral{2C}{\isacharcomma}}{\isaliteral{27}{\isacharprime}}nc{\isaliteral{29}{\isacharparenright}}abox{\isaliteral{5F}{\isacharunderscore}}impl\ list{\isaliteral{22}{\isachardoublequoteclose}}%
\begin{isamarkuptext}%
The type of abstraction of an (\emph {Abox\_impl}) to an \emph {Abox} is given by:%
\end{isamarkuptext}%
\isamarkuptrue%
\isacommand{type{\isaliteral{5F}{\isacharunderscore}}synonym}\isamarkupfalse%
\ {\isaliteral{28}{\isacharparenleft}}{\isaliteral{27}{\isacharprime}}ni{\isaliteral{2C}{\isacharcomma}}{\isaliteral{27}{\isacharprime}}nr{\isaliteral{2C}{\isacharcomma}}{\isaliteral{27}{\isacharprime}}nc{\isaliteral{29}{\isacharparenright}}abstraction{\isaliteral{3D}{\isacharequal}}{\isaliteral{22}{\isachardoublequoteopen}}{\isaliteral{28}{\isacharparenleft}}{\isaliteral{27}{\isacharprime}}ni{\isaliteral{2C}{\isacharcomma}}{\isaliteral{27}{\isacharprime}}nr{\isaliteral{2C}{\isacharcomma}}{\isaliteral{27}{\isacharprime}}nc{\isaliteral{29}{\isacharparenright}}abox{\isaliteral{5F}{\isacharunderscore}}impl{\isaliteral{5C3C52696768746172726F773E}{\isasymRightarrow}}{\isaliteral{28}{\isacharparenleft}}{\isaliteral{27}{\isacharprime}}ni{\isaliteral{2C}{\isacharcomma}}{\isaliteral{27}{\isacharprime}}nr{\isaliteral{2C}{\isacharcomma}}{\isaliteral{27}{\isacharprime}}nc{\isaliteral{29}{\isacharparenright}}abox{\isaliteral{22}{\isachardoublequoteclose}}%
\begin{isamarkuptext}%
Tableau is simply a list of \emph {Abox\_imp}.%
\end{isamarkuptext}%
\isamarkuptrue%
\isacommand{type{\isaliteral{5F}{\isacharunderscore}}synonym}\isamarkupfalse%
\ {\isaliteral{28}{\isacharparenleft}}{\isaliteral{27}{\isacharprime}}ni{\isaliteral{2C}{\isacharcomma}}{\isaliteral{27}{\isacharprime}}nr{\isaliteral{2C}{\isacharcomma}}{\isaliteral{27}{\isacharprime}}nc{\isaliteral{29}{\isacharparenright}}\ tableau\ \ {\isaliteral{3D}{\isacharequal}}\ {\isaliteral{22}{\isachardoublequoteopen}}{\isaliteral{28}{\isacharparenleft}}{\isaliteral{27}{\isacharprime}}ni{\isaliteral{2C}{\isacharcomma}}{\isaliteral{27}{\isacharprime}}nr{\isaliteral{2C}{\isacharcomma}}{\isaliteral{27}{\isacharprime}}nc{\isaliteral{29}{\isacharparenright}}\ abox{\isaliteral{5F}{\isacharunderscore}}impl\ list{\isaliteral{22}{\isachardoublequoteclose}}%
\isamarkupsubsection{Implementing Rules%
}
\isamarkuptrue%
\begin{isamarkuptext}%
Each rule is encoded as a pair consisting of the condition of applicability and the given action \emph {(Condition, Action)}. The type of condition is given by:%
\end{isamarkuptext}%
\isamarkuptrue%
\isacommand{type{\isaliteral{5F}{\isacharunderscore}}synonym}\isamarkupfalse%
\ {\isaliteral{28}{\isacharparenleft}}{\isaliteral{27}{\isacharprime}}ni{\isaliteral{2C}{\isacharcomma}}{\isaliteral{27}{\isacharprime}}nr{\isaliteral{2C}{\isacharcomma}}{\isaliteral{27}{\isacharprime}}nc{\isaliteral{29}{\isacharparenright}}appcond{\isaliteral{3D}{\isacharequal}}{\isaliteral{22}{\isachardoublequoteopen}}{\isaliteral{28}{\isacharparenleft}}{\isaliteral{27}{\isacharprime}}ni{\isaliteral{2C}{\isacharcomma}}{\isaliteral{27}{\isacharprime}}nr{\isaliteral{2C}{\isacharcomma}}{\isaliteral{27}{\isacharprime}}nc{\isaliteral{29}{\isacharparenright}}abox{\isaliteral{5F}{\isacharunderscore}}impl{\isaliteral{5C3C52696768746172726F773E}{\isasymRightarrow}}{\isaliteral{28}{\isacharparenleft}}{\isaliteral{27}{\isacharprime}}ni{\isaliteral{2C}{\isacharcomma}}{\isaliteral{27}{\isacharprime}}nr{\isaliteral{2C}{\isacharcomma}}{\isaliteral{27}{\isacharprime}}nc{\isaliteral{29}{\isacharparenright}}fact{\isaliteral{5C3C52696768746172726F773E}{\isasymRightarrow}}\ bool{\isaliteral{22}{\isachardoublequoteclose}}%
\begin{isamarkuptext}%
The type of the action of a rule is defined by:%
\end{isamarkuptext}%
\isamarkuptrue%
\isacommand{type{\isaliteral{5F}{\isacharunderscore}}synonym}\isamarkupfalse%
\ {\isaliteral{28}{\isacharparenleft}}{\isaliteral{27}{\isacharprime}}ni{\isaliteral{2C}{\isacharcomma}}{\isaliteral{27}{\isacharprime}}nr{\isaliteral{2C}{\isacharcomma}}{\isaliteral{27}{\isacharprime}}nc{\isaliteral{29}{\isacharparenright}}\ action\ {\isaliteral{3D}{\isacharequal}}\ \isanewline
\ \ \ \ {\isaliteral{22}{\isachardoublequoteopen}}{\isaliteral{28}{\isacharparenleft}}{\isaliteral{27}{\isacharprime}}ni{\isaliteral{2C}{\isacharcomma}}{\isaliteral{27}{\isacharprime}}nr{\isaliteral{2C}{\isacharcomma}}{\isaliteral{27}{\isacharprime}}nc{\isaliteral{29}{\isacharparenright}}abox{\isaliteral{5F}{\isacharunderscore}}impl\ {\isaliteral{2A}{\isacharasterisk}}\ {\isaliteral{28}{\isacharparenleft}}{\isaliteral{27}{\isacharprime}}ni{\isaliteral{2C}{\isacharcomma}}{\isaliteral{27}{\isacharprime}}nr{\isaliteral{2C}{\isacharcomma}}{\isaliteral{27}{\isacharprime}}nc{\isaliteral{29}{\isacharparenright}}fact\ {\isaliteral{2A}{\isacharasterisk}}\ {\isaliteral{28}{\isacharparenleft}}{\isaliteral{27}{\isacharprime}}ni{\isaliteral{2C}{\isacharcomma}}{\isaliteral{27}{\isacharprime}}nr{\isaliteral{2C}{\isacharcomma}}{\isaliteral{27}{\isacharprime}}nc{\isaliteral{29}{\isacharparenright}}abox{\isaliteral{5F}{\isacharunderscore}}impl\isanewline
\ \ \ \ \ \ \ \ \ \ \ \ \ \ \ \ \ \ \ \ \ \ \ \ \ \ \ \ \ \ \ \ \ \ \ \ \ \ \ \ \ \ \ \ \ \ \ \ \ \ \ \ {\isaliteral{5C3C52696768746172726F773E}{\isasymRightarrow}}{\isaliteral{28}{\isacharparenleft}}{\isaliteral{27}{\isacharprime}}ni{\isaliteral{2C}{\isacharcomma}}{\isaliteral{27}{\isacharprime}}nr{\isaliteral{2C}{\isacharcomma}}{\isaliteral{27}{\isacharprime}}nc{\isaliteral{29}{\isacharparenright}}abox{\isaliteral{5F}{\isacharunderscore}}impl\ list{\isaliteral{22}{\isachardoublequoteclose}}%
\begin{isamarkuptext}%
The rule is implemented in Isabelle by the function:%
\end{isamarkuptext}%
\isamarkuptrue%
\isacommand{datatype}\isamarkupfalse%
\ {\isaliteral{28}{\isacharparenleft}}{\isaliteral{27}{\isacharprime}}ni{\isaliteral{2C}{\isacharcomma}}{\isaliteral{27}{\isacharprime}}nr{\isaliteral{2C}{\isacharcomma}}{\isaliteral{27}{\isacharprime}}nc{\isaliteral{29}{\isacharparenright}}\ srule\ {\isaliteral{3D}{\isacharequal}}\ Rule\ {\isaliteral{22}{\isachardoublequoteopen}}{\isaliteral{28}{\isacharparenleft}}{\isaliteral{27}{\isacharprime}}ni{\isaliteral{2C}{\isacharcomma}}{\isaliteral{27}{\isacharprime}}nr{\isaliteral{2C}{\isacharcomma}}{\isaliteral{27}{\isacharprime}}nc{\isaliteral{29}{\isacharparenright}}appcond\ {\isaliteral{2A}{\isacharasterisk}}\ {\isaliteral{28}{\isacharparenleft}}{\isaliteral{27}{\isacharprime}}ni{\isaliteral{2C}{\isacharcomma}}{\isaliteral{27}{\isacharprime}}nr{\isaliteral{2C}{\isacharcomma}}{\isaliteral{27}{\isacharprime}}nc{\isaliteral{29}{\isacharparenright}}action{\isaliteral{22}{\isachardoublequoteclose}}%
\isadelimproof
\endisadelimproof
\isatagproof
\endisatagproof
{\isafoldproof}%
\isadelimproof
\endisadelimproof
\isadelimproof
\endisadelimproof
\isatagproof
\endisatagproof
{\isafoldproof}%
\isadelimproof
\endisadelimproof
\isadelimproof
\endisadelimproof
\isatagproof
\endisatagproof
{\isafoldproof}%
\isadelimproof
\endisadelimproof
\isadelimproof
\endisadelimproof
\isatagproof
\endisatagproof
{\isafoldproof}%
\isadelimproof
\endisadelimproof
\isadelimproof
\endisadelimproof
\isatagproof
\endisatagproof
{\isafoldproof}%
\isadelimproof
\endisadelimproof
\isadelimproof
\endisadelimproof
\isatagproof
\endisatagproof
{\isafoldproof}%
\isadelimproof
\endisadelimproof
\isadelimproof
\endisadelimproof
\isatagproof
\endisatagproof
{\isafoldproof}%
\isadelimproof
\endisadelimproof
\isadelimproof
\endisadelimproof
\isatagproof
\endisatagproof
{\isafoldproof}%
\isadelimproof
\endisadelimproof
\begin{isamarkuptext}%
Once the data structures and format rules are defined, we can encode each rule. For this, we determine for each rule the condition of its applicability and the action of this rule. For example the rule  $\rightarrow_{ \sqcap}$  is coded as follows:   
 The condition of applicability of the rule $\rightarrow_{ \sqcap}$ is given by the following function: 
\end{isamarkuptext}%
\isamarkuptrue%
\isacommand{fun}\isamarkupfalse%
\ appcond{\isaliteral{5F}{\isacharunderscore}}and\ {\isaliteral{3A}{\isacharcolon}}{\isaliteral{3A}{\isacharcolon}}\ {\isaliteral{22}{\isachardoublequoteopen}}{\isaliteral{28}{\isacharparenleft}}{\isaliteral{27}{\isacharprime}}ni{\isaliteral{2C}{\isacharcomma}}\ {\isaliteral{27}{\isacharprime}}nr{\isaliteral{2C}{\isacharcomma}}\ {\isaliteral{27}{\isacharprime}}nc{\isaliteral{29}{\isacharparenright}}\ appcond{\isaliteral{22}{\isachardoublequoteclose}}\ \isanewline
\ \ \ \ \isakeyword{where}\ {\isaliteral{22}{\isachardoublequoteopen}}appcond{\isaliteral{5F}{\isacharunderscore}}and\ Ab{\isaliteral{5F}{\isacharunderscore}}i\ {\isaliteral{28}{\isacharparenleft}}Inst\ x\ {\isaliteral{28}{\isacharparenleft}}AndC\ c{\isadigit{1}}\ c{\isadigit{2}}{\isaliteral{29}{\isacharparenright}}{\isaliteral{29}{\isacharparenright}}\ {\isaliteral{3D}{\isacharequal}}\ \isanewline
\ \ \ \ \ {\isaliteral{28}{\isacharparenleft}}{\isaliteral{5C3C6E6F743E}{\isasymnot}}{\isaliteral{28}{\isacharparenleft}}list{\isaliteral{5F}{\isacharunderscore}}ex\ {\isaliteral{28}{\isacharparenleft}}is{\isaliteral{5F}{\isacharunderscore}}x{\isaliteral{5F}{\isacharunderscore}}c{\isaliteral{5F}{\isacharunderscore}}inst\ x\ c{\isadigit{1}}{\isaliteral{29}{\isacharparenright}}\ Ab{\isaliteral{5F}{\isacharunderscore}}i{\isaliteral{29}{\isacharparenright}}\ {\isaliteral{5C3C6F723E}{\isasymor}}\ {\isaliteral{5C3C6E6F743E}{\isasymnot}}{\isaliteral{28}{\isacharparenleft}}list{\isaliteral{5F}{\isacharunderscore}}ex\ {\isaliteral{28}{\isacharparenleft}}is{\isaliteral{5F}{\isacharunderscore}}x{\isaliteral{5F}{\isacharunderscore}}c{\isaliteral{5F}{\isacharunderscore}}inst\ x\ c{\isadigit{2}}{\isaliteral{29}{\isacharparenright}}\ Ab{\isaliteral{5F}{\isacharunderscore}}i{\isaliteral{29}{\isacharparenright}}{\isaliteral{29}{\isacharparenright}}{\isaliteral{22}{\isachardoublequoteclose}}\isanewline
\ \ \ \ {\isaliteral{7C}{\isacharbar}}{\isaliteral{22}{\isachardoublequoteopen}}appcond{\isaliteral{5F}{\isacharunderscore}}and\ Ab{\isaliteral{5F}{\isacharunderscore}}i\ \ {\isaliteral{5F}{\isacharunderscore}}\ \ {\isaliteral{3D}{\isacharequal}}\ False{\isaliteral{22}{\isachardoublequoteclose}}%
\begin{isamarkuptext}%
The function \texttt{is\_x\_c\_inst} means: \isa{is{\isaliteral{5F}{\isacharunderscore}}x{\isaliteral{5F}{\isacharunderscore}}c{\isaliteral{5F}{\isacharunderscore}}inst\ x\ c\ f\ {\isaliteral{3D}{\isacharequal}}\ {\isaliteral{28}{\isacharparenleft}}f\ {\isaliteral{3D}{\isacharequal}}\ Inst\ x\ c{\isaliteral{29}{\isacharparenright}}}\\
The action provided by the application of this rule is:%
\end{isamarkuptext}%
\isamarkuptrue%
\isadelimproof
\endisadelimproof
\isatagproof
\endisatagproof
{\isafoldproof}%
\isadelimproof
\endisadelimproof
\isadelimproof
\endisadelimproof
\isatagproof
\endisatagproof
{\isafoldproof}%
\isadelimproof
\endisadelimproof
\isacommand{fun}\isamarkupfalse%
\ action{\isaliteral{5F}{\isacharunderscore}}and\ {\isaliteral{3A}{\isacharcolon}}{\isaliteral{3A}{\isacharcolon}}\ {\isaliteral{22}{\isachardoublequoteopen}}{\isaliteral{28}{\isacharparenleft}}{\isaliteral{27}{\isacharprime}}ni{\isaliteral{2C}{\isacharcomma}}{\isaliteral{27}{\isacharprime}}nr{\isaliteral{2C}{\isacharcomma}}{\isaliteral{27}{\isacharprime}}nc{\isaliteral{29}{\isacharparenright}}\ action{\isaliteral{22}{\isachardoublequoteclose}}\isanewline
\ \ \ \ \isakeyword{where}\ {\isaliteral{22}{\isachardoublequoteopen}}action{\isaliteral{5F}{\isacharunderscore}}and\ {\isaliteral{28}{\isacharparenleft}}prefix{\isaliteral{2C}{\isacharcomma}}\ {\isaliteral{28}{\isacharparenleft}}Inst\ x\ {\isaliteral{28}{\isacharparenleft}}AndC\ c{\isadigit{1}}\ c{\isadigit{2}}{\isaliteral{29}{\isacharparenright}}{\isaliteral{29}{\isacharparenright}}{\isaliteral{2C}{\isacharcomma}}\ suffix{\isaliteral{29}{\isacharparenright}}\ {\isaliteral{3D}{\isacharequal}}\ \isanewline
\ \ \ \ \ \ \ \ \ \ \ \ {\isaliteral{5B}{\isacharbrackleft}}{\isaliteral{5B}{\isacharbrackleft}}Inst\ x\ c{\isadigit{1}}{\isaliteral{2C}{\isacharcomma}}\ Inst\ x\ c{\isadigit{2}}{\isaliteral{5D}{\isacharbrackright}}\ {\isaliteral{40}{\isacharat}}\ prefix\ {\isaliteral{40}{\isacharat}}\ {\isaliteral{5B}{\isacharbrackleft}}Inst\ x\ {\isaliteral{28}{\isacharparenleft}}AndC\ c{\isadigit{1}}\ c{\isadigit{2}}{\isaliteral{29}{\isacharparenright}}{\isaliteral{5D}{\isacharbrackright}}\ {\isaliteral{40}{\isacharat}}\ suffix{\isaliteral{5D}{\isacharbrackright}}{\isaliteral{22}{\isachardoublequoteclose}}\isanewline
\ \ \ \ \ \ \ \ \ \ {\isaliteral{7C}{\isacharbar}}\ {\isaliteral{22}{\isachardoublequoteopen}}action{\isaliteral{5F}{\isacharunderscore}}and\ {\isaliteral{5F}{\isacharunderscore}}\ \ {\isaliteral{3D}{\isacharequal}}\ {\isaliteral{5B}{\isacharbrackleft}}{\isaliteral{5D}{\isacharbrackright}}{\isaliteral{22}{\isachardoublequoteclose}}%
\begin{isamarkuptext}%
\isa{prefix} denotes the elements of the list before the fact \isa{Inst\ x\ {\isaliteral{28}{\isacharparenleft}}AndC\ c{\isadigit{1}}\ c{\isadigit{2}}{\isaliteral{29}{\isacharparenright}}}  and \isa{suffix} denotes the elements  after the fact. The rule $\rightarrow_{ \sqcap}$is implemented by:%
\end{isamarkuptext}%
\isamarkuptrue%
\isacommand{definition}\isamarkupfalse%
\ and{\isaliteral{5F}{\isacharunderscore}}srule\ {\isaliteral{3A}{\isacharcolon}}{\isaliteral{3A}{\isacharcolon}}\ {\isaliteral{22}{\isachardoublequoteopen}}{\isaliteral{28}{\isacharparenleft}}{\isaliteral{27}{\isacharprime}}ni{\isaliteral{2C}{\isacharcomma}}{\isaliteral{27}{\isacharprime}}nr{\isaliteral{2C}{\isacharcomma}}{\isaliteral{27}{\isacharprime}}nc{\isaliteral{29}{\isacharparenright}}srule{\isaliteral{22}{\isachardoublequoteclose}}\ \isanewline
\ \ \ \ \ \ \ \isakeyword{where}\ {\isaliteral{22}{\isachardoublequoteopen}}and{\isaliteral{5F}{\isacharunderscore}}srule\ {\isaliteral{3D}{\isacharequal}}{\isaliteral{3D}{\isacharequal}}\ Rule\ {\isaliteral{28}{\isacharparenleft}}appcond{\isaliteral{5F}{\isacharunderscore}}and{\isaliteral{2C}{\isacharcomma}}action{\isaliteral{5F}{\isacharunderscore}}and{\isaliteral{29}{\isacharparenright}}{\isaliteral{22}{\isachardoublequoteclose}}%
\begin{isamarkuptext}%
Its application is defined in Isabelle by:%
\end{isamarkuptext}%
\isamarkuptrue%
\isacommand{definition}\isamarkupfalse%
\ and{\isaliteral{5F}{\isacharunderscore}}rule{\isaliteral{3A}{\isacharcolon}}{\isaliteral{3A}{\isacharcolon}}{\isaliteral{22}{\isachardoublequoteopen}}{\isaliteral{28}{\isacharparenleft}}{\isaliteral{27}{\isacharprime}}ni{\isaliteral{2C}{\isacharcomma}}{\isaliteral{27}{\isacharprime}}nr{\isaliteral{2C}{\isacharcomma}}{\isaliteral{27}{\isacharprime}}nc{\isaliteral{29}{\isacharparenright}}\ rule{\isaliteral{5F}{\isacharunderscore}}impl{\isaliteral{22}{\isachardoublequoteclose}}\ \isakeyword{where}\ {\isaliteral{22}{\isachardoublequoteopen}}and{\isaliteral{5F}{\isacharunderscore}}rule\ {\isaliteral{5C3C65717569763E}{\isasymequiv}}\ apply{\isaliteral{5F}{\isacharunderscore}}srule\ and{\isaliteral{5F}{\isacharunderscore}}srule{\isaliteral{22}{\isachardoublequoteclose}}%
\isadelimproof
\endisadelimproof
\isatagproof
\endisatagproof
{\isafoldproof}%
\isadelimproof
\endisadelimproof
\isadelimproof
\endisadelimproof
\isatagproof
\endisatagproof
{\isafoldproof}%
\isadelimproof
\endisadelimproof
\isadelimproof
\endisadelimproof
\isatagproof
\endisatagproof
{\isafoldproof}%
\isadelimproof
\endisadelimproof
\isadelimproof
\endisadelimproof
\isatagproof
\endisatagproof
{\isafoldproof}%
\isadelimproof
\endisadelimproof
\isadelimproof
\endisadelimproof
\isatagproof
\endisatagproof
{\isafoldproof}%
\isadelimproof
\endisadelimproof
\isadelimproof
\endisadelimproof
\isatagproof
\endisatagproof
{\isafoldproof}%
\isadelimproof
\endisadelimproof
\isadelimproof
\endisadelimproof
\isatagproof
\endisatagproof
{\isafoldproof}%
\isadelimproof
\endisadelimproof
\isadelimproof
\endisadelimproof
\isatagproof
\endisatagproof
{\isafoldproof}%
\isadelimproof
\endisadelimproof
\isadelimproof
\endisadelimproof
\isatagproof
\endisatagproof
{\isafoldproof}%
\isadelimproof
\endisadelimproof
\isadelimproof
\endisadelimproof
\isatagproof
\endisatagproof
{\isafoldproof}%
\isadelimproof
\endisadelimproof
\isadelimproof
\endisadelimproof
\isatagproof
\endisatagproof
{\isafoldproof}%
\isadelimproof
\endisadelimproof
\isadelimproof
\endisadelimproof
\isatagproof
\endisatagproof
{\isafoldproof}%
\isadelimproof
\endisadelimproof
\isadelimproof
\endisadelimproof
\isatagproof
\endisatagproof
{\isafoldproof}%
\isadelimproof
\endisadelimproof
\isadelimproof
\endisadelimproof
\isatagproof
\endisatagproof
{\isafoldproof}%
\isadelimproof
\endisadelimproof
\isadelimproof
\endisadelimproof
\isatagproof
\endisatagproof
{\isafoldproof}%
\isadelimproof
\endisadelimproof
\isadelimproof
\endisadelimproof
\isatagproof
\endisatagproof
{\isafoldproof}%
\isadelimproof
\endisadelimproof
\isadelimproof
\endisadelimproof
\isatagproof
\endisatagproof
{\isafoldproof}%
\isadelimproof
\endisadelimproof
\isadelimproof
\endisadelimproof
\isatagproof
\endisatagproof
{\isafoldproof}%
\isadelimproof
\endisadelimproof
\isadelimproof
\endisadelimproof
\isatagproof
\endisatagproof
{\isafoldproof}%
\isadelimproof
\endisadelimproof
\isadelimproof
\endisadelimproof
\isatagproof
\endisatagproof
{\isafoldproof}%
\isadelimproof
\endisadelimproof
\isadelimproof
\endisadelimproof
\isatagproof
\endisatagproof
{\isafoldproof}%
\isadelimproof
\endisadelimproof
\isadelimproof
\endisadelimproof
\isatagproof
\endisatagproof
{\isafoldproof}%
\isadelimproof
\endisadelimproof
\isadelimproof
\endisadelimproof
\isatagproof
\endisatagproof
{\isafoldproof}%
\isadelimproof
\endisadelimproof
\isadelimproof
\endisadelimproof
\isatagproof
\endisatagproof
{\isafoldproof}%
\isadelimproof
\endisadelimproof
\isadelimproof
\endisadelimproof
\isatagproof
\endisatagproof
{\isafoldproof}%
\isadelimproof
\endisadelimproof
\isadelimproof
\endisadelimproof
\isatagproof
\endisatagproof
{\isafoldproof}%
\isadelimproof
\endisadelimproof
\isadelimproof
\endisadelimproof
\isatagproof
\endisatagproof
{\isafoldproof}%
\isadelimproof
\endisadelimproof
\isadelimproof
\endisadelimproof
\isatagproof
\endisatagproof
{\isafoldproof}%
\isadelimproof
\endisadelimproof
\isadelimproof
\endisadelimproof
\isatagproof
\endisatagproof
{\isafoldproof}%
\isadelimproof
\endisadelimproof
\isadelimproof
\endisadelimproof
\isatagproof
\endisatagproof
{\isafoldproof}%
\isadelimproof
\endisadelimproof
\isadelimproof
\endisadelimproof
\isatagproof
\endisatagproof
{\isafoldproof}%
\isadelimproof
\endisadelimproof
\isadelimproof
\endisadelimproof
\isatagproof
\endisatagproof
{\isafoldproof}%
\isadelimproof
\endisadelimproof
\isadelimproof
\endisadelimproof
\isatagproof
\endisatagproof
{\isafoldproof}%
\isadelimproof
\endisadelimproof
\begin{isamarkuptext}%
To generalize our implementation for all $ \mathcal {ALC} $ rules, we define a list of implemented rules.%
\end{isamarkuptext}%
\isamarkuptrue%
\isacommand{definition}\isamarkupfalse%
\ list{\isaliteral{5F}{\isacharunderscore}}alc{\isaliteral{5F}{\isacharunderscore}}rules{\isaliteral{5F}{\isacharunderscore}}impl{\isaliteral{5F}{\isacharunderscore}}gen\ {\isaliteral{3A}{\isacharcolon}}{\isaliteral{3A}{\isacharcolon}}\isanewline
\ \ {\isaliteral{22}{\isachardoublequoteopen}}{\isaliteral{28}{\isacharparenleft}}{\isaliteral{28}{\isacharparenleft}}{\isaliteral{28}{\isacharparenleft}}{\isaliteral{27}{\isacharprime}}ni{\isaliteral{2C}{\isacharcomma}}{\isaliteral{27}{\isacharprime}}nr{\isaliteral{2C}{\isacharcomma}}{\isaliteral{27}{\isacharprime}}nc{\isaliteral{29}{\isacharparenright}}\ abox{\isaliteral{5F}{\isacharunderscore}}impl{\isaliteral{29}{\isacharparenright}}\ {\isaliteral{5C3C52696768746172726F773E}{\isasymRightarrow}}\ {\isaliteral{27}{\isacharprime}}ni{\isaliteral{29}{\isacharparenright}}\ {\isaliteral{5C3C52696768746172726F773E}{\isasymRightarrow}}\ {\isaliteral{28}{\isacharparenleft}}{\isaliteral{27}{\isacharprime}}ni{\isaliteral{3A}{\isacharcolon}}{\isaliteral{3A}{\isacharcolon}}alloc{\isaliteral{2C}{\isacharcomma}}\ {\isaliteral{27}{\isacharprime}}nr{\isaliteral{2C}{\isacharcomma}}\ {\isaliteral{27}{\isacharprime}}nc{\isaliteral{29}{\isacharparenright}}\ rule{\isaliteral{5F}{\isacharunderscore}}impl\ list{\isaliteral{22}{\isachardoublequoteclose}}\ \isakeyword{where}\isanewline
\ {\isaliteral{22}{\isachardoublequoteopen}}list{\isaliteral{5F}{\isacharunderscore}}alc{\isaliteral{5F}{\isacharunderscore}}rules{\isaliteral{5F}{\isacharunderscore}}impl{\isaliteral{5F}{\isacharunderscore}}gen\ gen\ {\isaliteral{3D}{\isacharequal}}{\isaliteral{3D}{\isacharequal}}\ \ {\isaliteral{5B}{\isacharbrackleft}}and{\isaliteral{5F}{\isacharunderscore}}rule{\isaliteral{2C}{\isacharcomma}}\ or{\isaliteral{5F}{\isacharunderscore}}rule{\isaliteral{2C}{\isacharcomma}}\ all{\isaliteral{5F}{\isacharunderscore}}rule{\isaliteral{2C}{\isacharcomma}}\ some{\isaliteral{5F}{\isacharunderscore}}rule{\isaliteral{5F}{\isacharunderscore}}gen\ gen{\isaliteral{5D}{\isacharbrackright}}{\isaliteral{22}{\isachardoublequoteclose}}%
\isamarkupsubsection{Soundness%
}
\isamarkuptrue%
\begin{isamarkuptext}%
We have shown in section  \ref{sec:correction}  the proof of soundness property of  rules on the abstract level. The following result allows us to demonstrate the soundness  of our implementation.

The idea can be illustrated by the following  diagram. Above, we see the proof on an abstract level: the application of an abstract rule (\isa{r{\isaliteral{5F}{\isacharunderscore}}logic}), applied to an \emph{Abox}, generates a set of \emph{Abox} successors. The implementation of \emph{Abox} by lists gives a \emph {Abox\_impl}, whose abstraction with a \isa{set} gives a set of facts. The application of a rule for this implementation (\isa{r{\isaliteral{5F}{\isacharunderscore}}impl})   must provide a list of  \emph{Abox} implementations that have the same abstraction. \input{adq} More formally, the definition in Isabelle:%
\end{isamarkuptext}%
\isamarkuptrue%
\isacommand{definition}\isamarkupfalse%
\ sound{\isaliteral{5F}{\isacharunderscore}}rule{\isaliteral{5F}{\isacharunderscore}}impl{\isaliteral{3A}{\isacharcolon}}{\isaliteral{3A}{\isacharcolon}}\isanewline
\ {\isaliteral{22}{\isachardoublequoteopen}}{\isaliteral{28}{\isacharparenleft}}{\isaliteral{27}{\isacharprime}}ni{\isaliteral{2C}{\isacharcomma}}{\isaliteral{27}{\isacharprime}}nr{\isaliteral{2C}{\isacharcomma}}{\isaliteral{27}{\isacharprime}}nc{\isaliteral{29}{\isacharparenright}}abstraction{\isaliteral{5C3C52696768746172726F773E}{\isasymRightarrow}}{\isaliteral{28}{\isacharparenleft}}{\isaliteral{27}{\isacharprime}}ni{\isaliteral{2C}{\isacharcomma}}{\isaliteral{27}{\isacharprime}}nr{\isaliteral{2C}{\isacharcomma}}{\isaliteral{27}{\isacharprime}}nc{\isaliteral{29}{\isacharparenright}}rule{\isaliteral{5C3C52696768746172726F773E}{\isasymRightarrow}}{\isaliteral{28}{\isacharparenleft}}{\isaliteral{27}{\isacharprime}}ni{\isaliteral{2C}{\isacharcomma}}{\isaliteral{27}{\isacharprime}}nr{\isaliteral{2C}{\isacharcomma}}{\isaliteral{27}{\isacharprime}}nc{\isaliteral{29}{\isacharparenright}}rule{\isaliteral{5F}{\isacharunderscore}}impl{\isaliteral{5C3C52696768746172726F773E}{\isasymRightarrow}}bool{\isaliteral{22}{\isachardoublequoteclose}}\ \isakeyword{where}\isanewline
{\isaliteral{22}{\isachardoublequoteopen}}sound{\isaliteral{5F}{\isacharunderscore}}rule{\isaliteral{5F}{\isacharunderscore}}impl\ abstr\ r\ r{\isaliteral{5F}{\isacharunderscore}}impl{\isaliteral{5C3C65717569763E}{\isasymequiv}}\ {\isaliteral{5C3C666F72616C6C3E}{\isasymforall}}\ ai\ ai{\isaliteral{27}{\isacharprime}}{\isaliteral{2E}{\isachardot}}{\isaliteral{28}{\isacharparenleft}}ai{\isaliteral{27}{\isacharprime}}\ {\isaliteral{5C3C696E3E}{\isasymin}}set{\isaliteral{28}{\isacharparenleft}}r{\isaliteral{5F}{\isacharunderscore}}impl\ ai{\isaliteral{29}{\isacharparenright}}{\isaliteral{29}{\isacharparenright}}{\isaliteral{5C3C6C6F6E6772696768746172726F773E}{\isasymlongrightarrow}}r{\isaliteral{28}{\isacharparenleft}}abstr\ ai{\isaliteral{29}{\isacharparenright}}{\isaliteral{28}{\isacharparenleft}}abstr\ ai{\isaliteral{27}{\isacharprime}}{\isaliteral{29}{\isacharparenright}}{\isaliteral{22}{\isachardoublequoteclose}}%
\isadelimproof
\endisadelimproof
\isatagproof
\endisatagproof
{\isafoldproof}%
\isadelimproof
\endisadelimproof
\isadelimproof
\endisadelimproof
\isatagproof
\endisatagproof
{\isafoldproof}%
\isadelimproof
\endisadelimproof
\isadelimproof
\endisadelimproof
\isatagproof
\endisatagproof
{\isafoldproof}%
\isadelimproof
\endisadelimproof
\isadelimproof
\endisadelimproof
\isatagproof
\endisatagproof
{\isafoldproof}%
\isadelimproof
\endisadelimproof
\isadelimproof
\endisadelimproof
\isatagproof
\endisatagproof
{\isafoldproof}%
\isadelimproof
\endisadelimproof
\isadelimproof
\endisadelimproof
\isatagproof
\endisatagproof
{\isafoldproof}%
\isadelimproof
\endisadelimproof
\isadelimproof
\endisadelimproof
\isatagproof
\endisatagproof
{\isafoldproof}%
\isadelimproof
\endisadelimproof
\isadelimproof
\endisadelimproof
\isatagproof
\endisatagproof
{\isafoldproof}%
\isadelimproof
\endisadelimproof
\isadelimproof
\endisadelimproof
\isatagproof
\endisatagproof
{\isafoldproof}%
\isadelimproof
\endisadelimproof
\isadelimproof
\endisadelimproof
\isatagproof
\endisatagproof
{\isafoldproof}%
\isadelimproof
\endisadelimproof
\isadelimproof
\endisadelimproof
\isatagproof
\endisatagproof
{\isafoldproof}%
\isadelimproof
\endisadelimproof
\isadelimproof
\endisadelimproof
\isatagproof
\endisatagproof
{\isafoldproof}%
\isadelimproof
\endisadelimproof
\isadelimproof
\endisadelimproof
\isatagproof
\endisatagproof
{\isafoldproof}%
\isadelimproof
\endisadelimproof
\isadelimproof
\endisadelimproof
\isatagproof
\endisatagproof
{\isafoldproof}%
\isadelimproof
\endisadelimproof
\isadelimproof
\endisadelimproof
\isatagproof
\endisatagproof
{\isafoldproof}%
\isadelimproof
\endisadelimproof
\isadelimproof
\endisadelimproof
\isatagproof
\endisatagproof
{\isafoldproof}%
\isadelimproof
\endisadelimproof
\isadelimproof
\endisadelimproof
\isatagproof
\endisatagproof
{\isafoldproof}%
\isadelimproof
\endisadelimproof
\isadelimproof
\endisadelimproof
\isatagproof
\endisatagproof
{\isafoldproof}%
\isadelimproof
\endisadelimproof
\isadelimproof
\endisadelimproof
\isatagproof
\endisatagproof
{\isafoldproof}%
\isadelimproof
\endisadelimproof
\isadelimproof
\endisadelimproof
\isatagproof
\endisatagproof
{\isafoldproof}%
\isadelimproof
\endisadelimproof
\isadelimproof
\endisadelimproof
\isatagproof
\endisatagproof
{\isafoldproof}%
\isadelimproof
\endisadelimproof
\isadelimproof
\endisadelimproof
\isatagproof
\endisatagproof
{\isafoldproof}%
\isadelimproof
\endisadelimproof
\isadelimproof
\endisadelimproof
\isatagproof
\endisatagproof
{\isafoldproof}%
\isadelimproof
\endisadelimproof
\isadelimproof
\endisadelimproof
\isatagproof
\endisatagproof
{\isafoldproof}%
\isadelimproof
\endisadelimproof
\isadelimproof
\endisadelimproof
\isatagproof
\endisatagproof
{\isafoldproof}%
\isadelimproof
\endisadelimproof
\isadelimproof
\endisadelimproof
\isatagproof
\endisatagproof
{\isafoldproof}%
\isadelimproof
\endisadelimproof
\isadelimproof
\endisadelimproof
\isatagproof
\endisatagproof
{\isafoldproof}%
\isadelimproof
\endisadelimproof
\isadelimproof
\endisadelimproof
\isatagproof
\endisatagproof
{\isafoldproof}%
\isadelimproof
\endisadelimproof
\isadelimproof
\endisadelimproof
\isatagproof
\endisatagproof
{\isafoldproof}%
\isadelimproof
\endisadelimproof
\isadelimproof
\endisadelimproof
\isatagproof
\endisatagproof
{\isafoldproof}%
\isadelimproof
\endisadelimproof
\isadelimproof
\endisadelimproof
\isatagproof
\endisatagproof
{\isafoldproof}%
\isadelimproof
\endisadelimproof
\isadelimproof
\endisadelimproof
\isatagproof
\endisatagproof
{\isafoldproof}%
\isadelimproof
\endisadelimproof
\isadelimproof
\endisadelimproof
\isatagproof
\endisatagproof
{\isafoldproof}%
\isadelimproof
\endisadelimproof
\isadelimproof
\endisadelimproof
\isatagproof
\endisatagproof
{\isafoldproof}%
\isadelimproof
\endisadelimproof
\isadelimproof
\endisadelimproof
\isatagproof
\endisatagproof
{\isafoldproof}%
\isadelimproof
\endisadelimproof
\isadelimproof
\endisadelimproof
\isatagproof
\endisatagproof
{\isafoldproof}%
\isadelimproof
\endisadelimproof
\isadelimproof
\endisadelimproof
\isatagproof
\endisatagproof
{\isafoldproof}%
\isadelimproof
\endisadelimproof
\isadelimproof
\endisadelimproof
\isatagproof
\endisatagproof
{\isafoldproof}%
\isadelimproof
\endisadelimproof
\isadelimproof
\endisadelimproof
\isatagproof
\endisatagproof
{\isafoldproof}%
\isadelimproof
\endisadelimproof
\isadelimproof
\endisadelimproof
\isatagproof
\endisatagproof
{\isafoldproof}%
\isadelimproof
\endisadelimproof
\isadelimproof
\endisadelimproof
\isatagproof
\endisatagproof
{\isafoldproof}%
\isadelimproof
\endisadelimproof
\isadelimproof
\endisadelimproof
\isatagproof
\endisatagproof
{\isafoldproof}%
\isadelimproof
\endisadelimproof
\isadelimproof
\endisadelimproof
\isatagproof
\endisatagproof
{\isafoldproof}%
\isadelimproof
\endisadelimproof
\isadelimproof
\endisadelimproof
\isatagproof
\endisatagproof
{\isafoldproof}%
\isadelimproof
\endisadelimproof
\isadelimproof
\endisadelimproof
\isatagproof
\endisatagproof
{\isafoldproof}%
\isadelimproof
\endisadelimproof
\isamarkupsubsection{Termination%
}
\isamarkuptrue%
\begin{isamarkuptext}%
Termination is an important property of rewriting systems. A standard method for proving termination of a rewrite system is to exhibit a well-founded ordering on terms, such that if $A_1$ is rewritten  to $A_2$  then  $A_1 \gg A_2$.

Formally, for proving termination, we associate to each implementation of a \emph {Abox} a measure. If this measure decreases with every rule in a well-founded ordering, termination is assured. In our case, a measure is a function of type $ Abox\_imp \rightarrow T$ for $T$ a domain equipped with a well founded relation $\gg $ in $T$ which we define in the sequel.%
\end{isamarkuptext}%
\isamarkuptrue%
\begin{isamarkuptext}%
We introduce the constructors necessary to define the measure:
 
The function \emph{sizeC} calculates the size of a concept, it is defined as the number of constructors of this concept.

\begin {itemize}
\item The measure in our case is a multiset of pairs of natural numbers.
\item For each axiom (\texttt{fact}) of \emph{Abox\_imp}, we associate a pair, depending on the structure of the axiom and the \emph {Abox \_imp}. 
\item The measure of the element that is applicable must decrease, without affecting the measure of other elements.
\end{itemize}
We now define  the measure of the axioms (this is the function \emph{meas\_comp}  defined in the following). If the axiom is:

\begin{itemize}
\item A relation $x \; r \; y$, we associate the pair $(0,0)$
\item An instance $x: D$. Depending on the structure of $ D $, there are three cases:
\end{itemize}

\begin{enumerate}
\item If $ D $ is an \emph{Atom} ($ x: A$) or \emph {Negation}
   ($x : \neg A$), we associate the value $(0,0)$ ;

\item If $ D $ is a conjunction, disjunction or existential quantifier: 
  \begin {itemize}
  \item If the corresponding rule is applicable on the axiom, we associate the pair $(sizeC(D), 0)$,
  \item else $(0,0)$;
  \end {itemize}

\item If $ D $ is a universal quantifier $(D = \forall r.
  C)$  we associate the pair $ (Comp_1, Comp_2) $ such that: 
\begin {enumerate}
  \item $Comp_1 = sizeC(D)$;
  \item $Comp_2 = Comp_{21} + Comp_{22}$ such that:
  \end {enumerate}
\end {enumerate}

\begin{itemize}
\item $ Comp_{21} $ is the number of applicability of the rule  $\rightarrow_ {\forall}$, ie the number of $x\;r\; y$ in \emph{Abox\_imp} such that $y : C$  is not in the \emph{Abox\_imp}. It is noted here that $Comp_{21}$  decreases if the rule $\rightarrow_{\forall}$  is applicable, but if we apply the rule $\rightarrow_{\exists}$  on another fact, this measure may increase. For this, we add the component $Comp_{22}$ which ensures that this remains constant by the application of another rule;
 
\item $Comp_{22}$ is the number of \emph{$\exists$-terms} reducible  and hidden (in the sense that they appear in the structure of the concept) in the \emph{Abox}. This value decreases if the rule $\rightarrow_{\exists}$ is applied and remains constant or decreases if another rule is applied.
\end{itemize}

In the end we can prove  that measure is well founded.%
\end{isamarkuptext}%
\isamarkuptrue%
\isacommand{lemma}\isamarkupfalse%
\ wf{\isaliteral{5F}{\isacharunderscore}}measure{\isaliteral{5F}{\isacharunderscore}}abox{\isaliteral{5F}{\isacharunderscore}}impl{\isaliteral{5F}{\isacharunderscore}}order{\isaliteral{3A}{\isacharcolon}}\ {\isaliteral{22}{\isachardoublequoteopen}}wf\ measure{\isaliteral{5F}{\isacharunderscore}}abox{\isaliteral{5F}{\isacharunderscore}}impl{\isaliteral{5F}{\isacharunderscore}}order{\isaliteral{22}{\isachardoublequoteclose}}%
\isadelimproof
\endisadelimproof
\isatagproof
\endisatagproof
{\isafoldproof}%
\isadelimproof
\endisadelimproof
\isadelimproof
\endisadelimproof
\isatagproof
\endisatagproof
{\isafoldproof}%
\isadelimproof
\endisadelimproof
\isadelimproof
\endisadelimproof
\isatagproof
\endisatagproof
{\isafoldproof}%
\isadelimproof
\endisadelimproof
\isadelimproof
\endisadelimproof
\isatagproof
\endisatagproof
{\isafoldproof}%
\isadelimproof
\endisadelimproof
\isadelimproof
\endisadelimproof
\isatagproof
\endisatagproof
{\isafoldproof}%
\isadelimproof
\endisadelimproof
\isadelimproof
\endisadelimproof
\isatagproof
\endisatagproof
{\isafoldproof}%
\isadelimproof
\endisadelimproof
\isadelimproof
\endisadelimproof
\isatagproof
\endisatagproof
{\isafoldproof}%
\isadelimproof
\endisadelimproof
\isadelimproof
\endisadelimproof
\isatagproof
\endisatagproof
{\isafoldproof}%
\isadelimproof
\endisadelimproof
\isadelimproof
\endisadelimproof
\isatagproof
\endisatagproof
{\isafoldproof}%
\isadelimproof
\endisadelimproof
\isadelimtheory
\endisadelimtheory
\isatagtheory
\endisatagtheory
{\isafoldtheory}%
\isadelimtheory
\endisadelimtheory
\end{isabellebody}%

%% file: adq.tex
\begin{center}
\mbox{
\xymatrix@=15mm{
     & \emph{Abox} \ar@{->}[r]_{r\_logic} &      \emph{set of Abox}    & \\
     & \emph{Abox\_impl} \ar@{->}[u]^{Abstraction (set)} \ar@{->}[r]^{r\_impl}       &    \emph{tableau} \ar@{->}[u]   & 
  }}
\end{center}


%% file: conclusion.tex
\section{Conclusion}\label{sec:conclusion}

In this paper we have presented a definition of a reasoner validated for the description logic $ \mathcal {ALC} $ based on the method of semantic tableaux. This formalization in Isabelle and development is based on several modules:
 
\begin{itemize}
\item Specifying the syntax and semantics of $ \cal {ALC} $,
\item Coding of \emph{ Abox} and the formalization of the transformation rules of semantic tableaux,
\item The proof of the soundness and completeness of semantic tableaux,

\item The proof of termination requires the definition of a   measure for each \emph {Abox}. We have shown that this measure decreases for each application of a rule.
\item The implementation of \emph{Abox}, tableaux and rules transformation 
\item Defining a strategy of proof for $ \mathcal {ALC} $
\item Finally, the extraction of an executable, certified reasoner in the Caml language.
\end {itemize}
We will consider several extensions of this work, among which are:
\begin {itemize}
\item Extensions of this work  to more expressive logics used in the semantic Web, such as $ \mathcal {SHOIQ} $ and $\mathcal{SHOIN}$.
\item Sets of refinements by other more efficient data structures as lists, and in particular the use of indexing techniques to speed up testing unsatisfiability of a table (`` clash'') or to identify the applicable rules.

\end {itemize}


%% file: main.bbl
\begin{thebibliography}{10}
\providecommand{\bibitemdeclare}[2]{}
\providecommand{\surnamestart}{}
\providecommand{\surnameend}{}
\providecommand{\urlprefix}{Available at }
\providecommand{\url}[1]{\texttt{#1}}
\providecommand{\href}[2]{\texttt{#2}}
\providecommand{\urlalt}[2]{\href{#1}{#2}}
\providecommand{\doi}[1]{doi:\urlalt{http://dx.doi.org/#1}{#1}}
\providecommand{\bibinfo}[2]{#2}

\bibitemdeclare{book}{dlhandbook}
\bibitem{dlhandbook}
\bibinfo{author}{Franz \surnamestart Baader\surnameend}, \bibinfo{author}{Diego
  \surnamestart Calvanese\surnameend}, \bibinfo{author}{Deborah~L.
  \surnamestart McGuinness\surnameend}, \bibinfo{author}{Daniele \surnamestart
  Nardi\surnameend} \& \bibinfo{author}{Peter~F. \surnamestart
  Patel-Schneider\surnameend} (\bibinfo{year}{2007}): \emph{\bibinfo{title}{The
  Description Logic Handbook: Theory, Implementation, and Applications}}.
\newblock \bibinfo{publisher}{Cambridge University Press},
  \doi{10.1017/CBO9780511711787}.

\bibitemdeclare{inproceedings}{Baader1991}
\bibitem{Baader1991}
\bibinfo{author}{Franz \surnamestart Baader\surnameend} \&
  \bibinfo{author}{Philipp \surnamestart Hanschke\surnameend}
  (\bibinfo{year}{1991}): \emph{\bibinfo{title}{A schema for integrating
  concrete domains into concept languages}}.
\newblock In: {\sl \bibinfo{booktitle}{Proc. of the 12th Int. Joint Conf. on
  Artificial Intelligence (IJCAI'91)}}, pp. \bibinfo{pages}{452--457}.

\bibitemdeclare{inproceedings}{BaaderHS05}
\bibitem{BaaderHS05}
\bibinfo{author}{Franz \surnamestart Baader\surnameend}, \bibinfo{author}{Ian
  \surnamestart Horrocks\surnameend} \& \bibinfo{author}{Ulrike \surnamestart
  Sattler\surnameend} (\bibinfo{year}{2005}): \emph{\bibinfo{title}{Description
  Logics as Ontology Languages for the Semantic Web}}.
\newblock In: {\sl \bibinfo{booktitle}{Mechanizing Mathematical Reasoning}},
  pp. \bibinfo{pages}{228--248}, \doi{10.1007/978-3-540-32254-2\_14}.
\newblock
  \urlprefix\url{http://www.informatik.uni-trier.de/~ley/db/conf/birthday/siek%
mann2005.html#BaaderHS05}.

\bibitemdeclare{article}{BaaderSattler-JLC-99}
\bibitem{BaaderSattler-JLC-99}
\bibinfo{author}{Franz \surnamestart Baader\surnameend} \&
  \bibinfo{author}{Ulrike \surnamestart Sattler\surnameend}
  (\bibinfo{year}{1999}): \emph{\bibinfo{title}{Expressive Number Restrictions
  in Description Logics}}.
\newblock {\sl \bibinfo{journal}{Journal of Logic and Computation}}
  \bibinfo{volume}{9}(\bibinfo{number}{3}), pp. \bibinfo{pages}{319--350},
  \doi{10.1093/logcom/9.3.319}.
\newblock
  \urlprefix\url{http://lat.inf.tu-dresden.de/research/papers/1999/BaaderSattl%
er-JLC-99.ps.gz}.

\bibitemdeclare{article}{Baader00tableaualgorithms}
\bibitem{Baader00tableaualgorithms}
\bibinfo{author}{Franz \surnamestart Baader\surnameend} \&
  \bibinfo{author}{Ulrike \surnamestart Sattler\surnameend}
  (\bibinfo{year}{2001}): \emph{\bibinfo{title}{An Overview of Tableau
  Algorithms for Description Logics}}.
\newblock {\sl \bibinfo{journal}{Studia Logica}}
  \bibinfo{volume}{69}(\bibinfo{number}{1}), pp. \bibinfo{pages}{5--40},
  \doi{10.1023/A:1013882326814}.
\newblock
  \urlprefix\url{http://lat.inf.tu-dresden.de/research/papers/2001/BaaderSattl%
er-StudiaLogica.ps.gz}.

\bibitemdeclare{inproceedings}{Berardi01reasoningon}
\bibitem{Berardi01reasoningon}
\bibinfo{author}{Daniela \surnamestart Berardi\surnameend},
  \bibinfo{author}{Diego \surnamestart Calvanese\surnameend} \&
  \bibinfo{author}{Guiseppe~De \surnamestart Giacomo\surnameend}
  (\bibinfo{year}{2001}): \emph{\bibinfo{title}{Reasoning on {UML} Class
  Diagrams using Description Logic Based Systems}}.
\newblock In: {\sl \bibinfo{booktitle}{Proc of the KI 2001 Workshop on
  Applications of Description Logics, CEUR Electronic Workshop Proceedings,
  http://ceur-ws.org/Vol-44}}.

\bibitemdeclare{inproceedings}{chaabani09:_formal_de_la_logiq_descr}
\bibitem{chaabani09:_formal_de_la_logiq_descr}
\bibinfo{author}{Mohamed \surnamestart Chaabani\surnameend},
  \bibinfo{author}{Mohamed \surnamestart Mezghiche\surnameend} \&
  \bibinfo{author}{Martin \surnamestart Strecker\surnameend}
  (\bibinfo{year}{2009}): \emph{\bibinfo{title}{Formalisation de la logique de
  description $\mathcal {ALC}$ dans l'assistant de preuve {Coq}}}.
\newblock In \bibinfo{editor}{L.~\surnamestart Bellatrache\surnameend},
  \bibinfo{editor}{G.~\surnamestart Kassel\surnameend} \&
  \bibinfo{editor}{P.~\surnamestart Thiran\surnameend}, editors: {\sl
  \bibinfo{booktitle}{Proc. 3es Journ{\'e}es francophones sur les ontologies}},
  pp. \bibinfo{pages}{139--147}.

\bibitemdeclare{inproceedings}{chaabani10_afadl}
\bibitem{chaabani10_afadl}
\bibinfo{author}{Mohamed \surnamestart Chaabani\surnameend},
  \bibinfo{author}{Mohamed \surnamestart Mezghiche\surnameend} \&
  \bibinfo{author}{Martin \surnamestart Strecker\surnameend}
  (\bibinfo{year}{2010}): \emph{\bibinfo{title}{V{\'e}rification d'une
  m{\'e}thode de preuve pour la logique de description {$\cal{ALC}$}}}.
\newblock In: {\sl \bibinfo{booktitle}{Proc. 10{\`e}me Journ{\'e}es Approches
  Formelles dans l'Assistance au D{\'e}veloppement de Logiciels}}, pp.
  \bibinfo{pages}{149--163}.

\bibitemdeclare{inproceedings}{Fehreretal94a}
\bibitem{Fehreretal94a}
\bibinfo{author}{Detlef \surnamestart Fehrer\surnameend},
  \bibinfo{author}{Ullrich \surnamestart Hustadt\surnameend},
  \bibinfo{author}{Manfred \surnamestart Jaeger\surnameend},
  \bibinfo{author}{Andreas \surnamestart Nonnengart\surnameend},
  \bibinfo{author}{Hans~J{\"u}rgen \surnamestart Ohlbach\surnameend},
  \bibinfo{author}{Renate~A. \surnamestart Schmidt\surnameend},
  \bibinfo{author}{Christoph \surnamestart Weidenbach\surnameend} \&
  \bibinfo{author}{Emil \surnamestart Weydert\surnameend}
  (\bibinfo{year}{1994}): \emph{\bibinfo{title}{Description Logics for Natural
  Language Processing}}.
\newblock In \bibinfo{editor}{Franz \surnamestart Baader\surnameend},
  \bibinfo{editor}{Maurizio \surnamestart Lenzerini\surnameend},
  \bibinfo{editor}{Werner \surnamestart Nutt\surnameend} \&
  \bibinfo{editor}{Peter~F. \surnamestart Patel-Schneider\surnameend}, editors:
  {\sl \bibinfo{booktitle}{International Workshop on Description Logics '94}},
  {\sl \bibinfo{series}{Document}} \bibinfo{volume}{D-94-10},
  \bibinfo{publisher}{DFKI}, \bibinfo{address}{Bonn, Germany}, pp.
  \bibinfo{pages}{80--84}.

\bibitemdeclare{inproceedings}{Racer}
\bibitem{Racer}
\bibinfo{author}{Volker \surnamestart Haarslev\surnameend} \&
  \bibinfo{author}{Ralf \surnamestart M\"{o}ller\surnameend}
  (\bibinfo{year}{2001}): \emph{\bibinfo{title}{RACER System Description}}.
\newblock In: {\sl \bibinfo{booktitle}{IJCAR '01: Proc. First International
  Joint Conference on Automated Reasoning}},
  \bibinfo{publisher}{Springer-Verlag}, pp. \bibinfo{pages}{701--706},
  \doi{10.1007/3-540-45744-5\_59}.

\bibitemdeclare{inproceedings}{hidalgo07:_alc}
\bibitem{hidalgo07:_alc}
\bibinfo{author}{Mar{\'i}a-Jos{\'e} \surnamestart Hidalgo\surnameend},
  \bibinfo{author}{Jos{\'e}-Antonio \surnamestart Alonso\surnameend},
  \bibinfo{author}{Joaqu{\'i}n \surnamestart Borrego-D{\'i}az\surnameend},
  \bibinfo{author}{Francisco-Jesus \surnamestart Martin-Mateos\surnameend} \&
  \bibinfo{author}{Jos{\'e}-Luis \surnamestart Ruiz-Reina\surnameend}
  (\bibinfo{year}{2007}): \emph{\bibinfo{title}{A formally verified prover for
  the ALC description logic.}}
\newblock In: {\sl \bibinfo{booktitle}{20th International Conference on Theorem
  Proving in Higher Order Logics, TPHOLs 2007}}, {\sl \bibinfo{series}{Lecture
  Notes in Computer Science}} \bibinfo{volume}{4732}, pp.
  \bibinfo{pages}{135--150}, \doi{10.1007/978-3-540-74591-4\_11}.
\newblock \urlprefix\url{http://www.cs.us.es/~mjoseh/pub/2007-TPHOLs.pdf}.

\bibitemdeclare{article}{Horrocks}
\bibitem{Horrocks}
\bibinfo{author}{Ian \surnamestart Horrocks\surnameend} \&
  \bibinfo{author}{Ulrike \surnamestart Sattler\surnameend}
  (\bibinfo{year}{2007}): \emph{\bibinfo{title}{A Tableau Decision Procedure
  for $\mathcal{SHOIQ}$}}.
\newblock {\sl \bibinfo{journal}{J.\ of Automated Reasoning}}
  \bibinfo{volume}{39}(\bibinfo{number}{3}), pp. \bibinfo{pages}{249--276},
  \doi{10.1007/s10817-007-9079-9}.
\newblock \urlprefix\url{download/2007/HoSa07a.pdf}.

\bibitemdeclare{inproceedings}{luther09:_who_heck_father_bob}
\bibitem{luther09:_who_heck_father_bob}
\bibinfo{author}{Marko \surnamestart Luther\surnameend},
  \bibinfo{author}{Thorsten \surnamestart Liebig\surnameend},
  \bibinfo{author}{Sebastian \surnamestart B{\"o}hm\surnameend} \&
  \bibinfo{author}{Olaf \surnamestart Noppens\surnameend}
  (\bibinfo{year}{2009}): \emph{\bibinfo{title}{{Who the Heck is the Father of
  Bob?}}}
\newblock In: {\sl \bibinfo{booktitle}{6th Annual European Semantic Web
  Conference (ESWC2009)}}, pp. \bibinfo{pages}{66--80}.
\newblock
  \urlprefix\url{http://data.semanticweb.org/conference/eswc/2009/paper/222}.

\bibitemdeclare{book}{Isabelle_Tutorial}
\bibitem{Isabelle_Tutorial}
\bibinfo{author}{Tobias \surnamestart Nipkow\surnameend},
  \bibinfo{author}{Lawrence \surnamestart Paulson\surnameend} \&
  \bibinfo{author}{Markus \surnamestart Wenzel\surnameend}
  (\bibinfo{year}{2002}): \emph{\bibinfo{title}{Isabelle/HOL. A Proof Assistant
  for Higher-Order Logic}}.
\newblock {\sl \bibinfo{series}{Lecture Notes in Computer Science}}
  \bibinfo{volume}{2283}, \bibinfo{publisher}{Springer},
  \doi{10.1007/3-540-45949-9}.

\bibitemdeclare{article}{DBLP:conf/kr/DoniniLNN91}
\bibitem{DBLP:conf/kr/DoniniLNN91}
\bibinfo{author}{Werner \surnamestart Nutt\surnameend},
  \bibinfo{author}{Francesco~M. \surnamestart Donini\surnameend},
  \bibinfo{author}{Maurizio \surnamestart Lenzerini\surnameend} \&
  \bibinfo{author}{Daniele \surnamestart Nardi\surnameend}
  (\bibinfo{year}{1997}): \emph{\bibinfo{title}{The complexity of concept
  languages}}.
\newblock {\sl \bibinfo{journal}{Inf. Comput.}}
  \bibinfo{volume}{134}(\bibinfo{number}{1}), pp. \bibinfo{pages}{1--58},
  \doi{10.1006/inco.1997.2625}.

\bibitemdeclare{inproceedings}{ridge05:_fol}
\bibitem{ridge05:_fol}
\bibinfo{author}{Tom \surnamestart Ridge\surnameend} \& \bibinfo{author}{James
  \surnamestart Margetson\surnameend} (\bibinfo{year}{2005}):
  \emph{\bibinfo{title}{A mechanically verified, sound and complete theorem
  prover for {FOL}}}.
\newblock In \bibinfo{editor}{Joe \surnamestart Hurd\surnameend} \&
  \bibinfo{editor}{Tom \surnamestart Melham\surnameend}, editors: {\sl
  \bibinfo{booktitle}{18th International Conference on Theorem Proving in
  Higher Order Logics: TPHOLs 2005}}, {\sl \bibinfo{series}{Lecture Notes in
  Computer Science}} \bibinfo{volume}{3603}, \doi{10.1007/11541868\_19}.

\bibitemdeclare{inproceedings}{schimpf:construction}
\bibitem{schimpf:construction}
\bibinfo{author}{Alexander \surnamestart Schimpf\surnameend},
  \bibinfo{author}{Stephan \surnamestart Merz\surnameend} \&
  \bibinfo{author}{Jan-Georg \surnamestart Smaus\surnameend}
  (\bibinfo{year}{2009}): \emph{\bibinfo{title}{Construction of {B\"uchi}
  Automata for {LTL} Model Checking Verified in {I}sabelle/{HOL}}}.
\newblock In \bibinfo{editor}{Tobias \surnamestart Nipkow\surnameend} \&
  \bibinfo{editor}{Christian \surnamestart Urban\surnameend}, editors: {\sl
  \bibinfo{booktitle}{22nd Intl. Conf. Theorem Proving in Higher-Order Logics
  (TPHOLs 2009)}}, {\sl \bibinfo{series}{Lecture Notes in Computer Science}}
  \bibinfo{volume}{5674}, \bibinfo{publisher}{Springer},
  \bibinfo{address}{Munich, Germany}, \doi{10.1007/978-3-642-03359-9\_29}.
\newblock \urlprefix\url{http://www.loria.fr/~merz/papers/tphols2009.html}.

\bibitemdeclare{article}{Schmidt-SchaubB:1991:ACD:114341.114342}
\bibitem{Schmidt-SchaubB:1991:ACD:114341.114342}
\bibinfo{author}{Manfred \surnamestart Schmidt-Schaub\ss\surnameend} \&
  \bibinfo{author}{Gert \surnamestart Smolka\surnameend}
  (\bibinfo{year}{1991}): \emph{\bibinfo{title}{Attributive concept
  descriptions with complements}}.
\newblock {\sl \bibinfo{journal}{Artif. Intell.}}
  \bibinfo{volume}{48}(\bibinfo{number}{1}), pp. \bibinfo{pages}{1--26},
  \doi{10.1016/0004-3702(91)90078-X}.

\bibitemdeclare{inproceedings}{Tsarkov2006}
\bibitem{Tsarkov2006}
\bibinfo{author}{Dmitry \surnamestart Tsarkov\surnameend} \&
  \bibinfo{author}{Ian \surnamestart Horrocks\surnameend}
  (\bibinfo{year}{2006}): \emph{\bibinfo{title}{FaCT++ Description Logic
  Reasoner: System Description}}.
\newblock In: {\sl \bibinfo{booktitle}{Proc. of the Int. Joint Conf. on
  Automated Reasoning (IJCAR 2006)}}, {\sl \bibinfo{series}{Lecture Notes in
  Artificial Intelligence}} \bibinfo{volume}{4130},
  \bibinfo{publisher}{Springer}, pp. \bibinfo{pages}{292--297},
  \doi{10.1007/11814771\_26}.

\bibitemdeclare{mastersthesis}{wind01:_modal_logic}
\bibitem{wind01:_modal_logic}
\bibinfo{author}{Paulien \surnamestart de~Wind\surnameend}
  (\bibinfo{year}{2001}): \emph{\bibinfo{title}{Modal Logic}}.
\newblock Master's thesis, \bibinfo{school}{Vrije Universiteit Amsterdam}.

\end{thebibliography}
